\documentclass{emulateapj}

\usepackage{graphicx}
\usepackage{bm}
\usepackage{amssymb}
\usepackage{amsmath}
\usepackage{cancel}

\usepackage{epstopdf}
\usepackage{natbib}
\usepackage{epsfig}

\newcommand{\avg}[1]{\ensuremath{\langle #1 \rangle}}
\newcommand{\bma}{\begin{math}}
\newcommand{\ema}{\end{math}}
\newcommand{\beq}{\begin{equation}}
\newcommand{\eeq}{\end{equation}}
\newcommand{\beqa}{\begin{eqnarray}}
\newcommand{\eeqa}{\end{eqnarray}}
\newcommand{\bc}{\begin{center}}
\newcommand{\ec}{\end{center}} 
\newcommand{\bit}{\begin{itemize}}
\newcommand{\eit}{\end{itemize}}

\font\BFd=cmmib10
\font\BFt=cmmib10
\font\BFs=cmmib10 scaled 700
\font\BFss=cmmib10 scaled 500

\def\bbox#1{%
\relax\ifmmode
\mathchoice
{{\hbox{\BFd #1}}}
{{\hbox{\BFt #1}}}
{{\hbox{\BFs #1}}}
{{\hbox{\BFss #1}}}
\else \mbox{#1} \fi }

\def\x{{\bbox{x}}}

\begin{document}

\submitted{\today. To be submitted to \apj.} 

\title{On Modeling and Measuring the Temperature of the $z \sim 5$ IGM}
\author{Adam Lidz\altaffilmark{1} \& Matthew Malloy\altaffilmark{1}}
\altaffiltext{1} {Department of Physics \& Astronomy, University of Pennsylvania, 209 South 33rd Street, Philadelphia, PA 19104, USA}
\email{alidz@sas.upenn.edu}

\begin{abstract}
The temperature of the low-density intergalactic medium (IGM) at high redshift is sensitive to the timing and nature of
hydrogen and HeII reionization, and can be measured from Lyman-alpha (Ly-$\alpha$) forest absorption spectra.
Since the memory of intergalactic gas to heating during reionization gradually fades, measurements
as close as possible to reionization are desirable. In addition, measuring the IGM temperature at sufficiently 
high redshifts should help to isolate
the effects of hydrogen reionization since HeII reionization starts later, at lower redshift.
Motivated by this, we model the IGM temperature at $z \gtrsim 5$ using semi-numeric models of patchy reionization.  
We construct
mock Ly-$\alpha$ forest spectra from these models and consider their observable implications. We find that the small-scale
structure in the Ly-$\alpha$ forest is sensitive to the temperature of the IGM even at redshifts where the average
absorption in the forest is as high as $90\%$. We forecast the accuracy at which the $z \gtrsim 5$ IGM temperature
can be measured using existing samples of high resolution quasar spectra, and find that interesting constraints are
possible. For example, an early reionization model
in which reionization ends at $z \sim 10$ should be distinguishable -- at high statistical significance -- from a lower redshift model where
reionization completes at $z \sim 6$. 
We discuss
improvements to our modeling that may be required to robustly interpret future measurements.
\end{abstract}


\section{Introduction} \label{sec:intro}

The temperature of the low density intergalactic medium (IGM) after reionization retains information about
when and how the gas was heated during the Epoch of Reionization (EoR) (e.g. \citealt{1994MNRAS.266..343M,Hui:1997dp,Theuns:2002yc,Hui:2003hn}).
The temperature of the IGM in turn
impacts the statistical properties of the Ly-$\alpha$ forest towards background quasars and so the absorption in the
forest provides ``fossil'' evidence regarding the timing and nature of reionization. Scrutinized carefully, this
fossil may therefore improve our understanding of reionization. For example, the IGM will likely be cooler at $z \sim 5$ if most of the IGM volume
reionized at relatively high redshift, near e.g. $z \sim 10$, than if reionization happened later, near say $z \sim 6$. If reionization occurs
early, the gas has longer to cool and reaches a lower temperature than if it happens late, at least provided the gas is heated to a fixed
temperature at reionization. In addition, the IGM temperature should be inhomogeneous, partly as a result of spatial
variations in the timing of reionization across the universe \citep{Trac:2008yz,Cen:2009bg,Furlanetto:2009kr}. Careful measurements of the IGM temperature after reionization should hence
constrain the average reionization history of the universe, and may potentially reveal spatial variations around the average 
history as well.

Two separate phases of reionization are likely relevant for understanding the thermal history of the IGM: an early period
of hydrogen reionization during which hydrogen is ionized, and helium is singly ionized by star-forming galaxies, and a later period 
in which helium is doubly-ionized by quasars, i.e. HeII reionization. Hydrogen reionization completed sometime 
before $z \sim 6$ or so (e.g. \citealt{Fan:2005es}, although it might conceivably end
as late as $z \sim 5$ -- see \citealt{McGreer:2011dm,Mesinger:2009mv,Lidz:2007mz}) ,
while mounting evidence suggests HeII reionization finished by $z \gtrsim 2.5-3$ (see e.g. \citealt{Worseck:2011qk,Syphers:2011uw} and references therein). 
Many of the existing IGM temperature measurements focus on redshifts of
$z \sim 2-4$ \citep{Schaye:1999vr,Ricotti:1999hx,McDonald:2000nn,Zaldarriaga:2000mz,Theuns:2001my,Lidz:2009ca};
in this case the temperature is likely strongly influenced by HeII
reionization (e.g. \citealt{McQuinn:2008am}) and so these measurements
mostly constrain helium reionization rather than hydrogen reionization.

In order to best constrain hydrogen reionization using the
thermal history of the IGM, temperature measurements
at higher redshift ($z \gtrsim 5$) are required. Indeed, recent work
has started to probe the temperature at these early times. In particular,
the recent study by \citet{Becker:2012aq} includes a measurement
at $z=4.8$; \citet{Bolton:2011ck} and \citet{Raskutti:2012qz} determine
the $z \sim 6$ temperature in the special ``proximity zone'' region of the Ly-$\alpha$
forest close to the quasar itself; and the analysis in \citet{Viel:2013fqw} starts to bound
the $z \gtrsim 5$ IGM temperature, although these authors focus on placing
limits on warm dark matter models.

The temperature at these higher redshifts is unlikely to be significantly impacted by HeII reionization. 
In addition, the ``memory'' of intergalactic gas to heating during the EoR gradually fades and so measurements as close as possible
to the EoR should, in principle, be most constraining.
It is not, however, obvious that the IGM temperature can be measured accurately enough 
from the $z \gtrsim 5$ Ly-$\alpha$ forest to exploit the sensitivity of the high redshift temperature to the properties of reionization. In particular, the forest is highly absorbed by $z \sim 5$ with $z \gtrsim 6$ spectra showing
essentially complete Gunn-Peterson \citep{1965ApJ...142.1633G} absorption troughs \citep{Becker:2001ee,Fan:2005es}. An interesting question is then: what is the highest redshift at which it is feasible
to measure the IGM temperature from the Ly-$\alpha$ forest?

Towards this end, the goal of this paper is to both model the thermal state of the $z \sim 5$ IGM, incorporating inhomogeneities in
the hydrogen reionization process, and to quantify the prospects for actually measuring the IGM temperature using $z \gtrsim 5$ Ly-$\alpha$ forest absorption spectra. The outline of this paper is as follows. In \S \ref{sec:sims}, we describe
the numerical simulations used in our analysis. In \S \ref{sec:reion_hist}, we present plausible example models for
the reionization history of the universe and describe our approach for modeling inhomogeneous reionization. We adopt a semi-analytic approach
for modeling the resulting thermal history of the IGM, as described in \S \ref{sec:therm_hist}. In this section, we also quantify
the statistical properties of the temperature field in several simulated reionization models.
 Finally, in \S \ref{sec:temp_measure} we discuss how to measure
the temperature from the $z \sim 5$ Ly-$\alpha$ forest, and forecast how well it may be measured with existing data. Our main conclusions
are described in \S \ref{sec:conclusions}.

This work partly overlaps with previous work which also recognized the importance of, and modeled, temperature inhomogeneities in the $z \sim 5$ IGM
and considered some of the observable implications \citep{Trac:2008yz,Cen:2009bg,Furlanetto:2009kr}.\footnote{\citet{Lai:2005ha} also considered
temperature fluctuations from hydrogen reionization, but these authors focused on $z \sim 3$ where -- as they discussed -- these fluctuations should be small and swamped
by effects from HeII reionization.} One key difference with this earlier work is
that we consider a more direct approach for measuring the temperature of the $z \sim 5$ IGM from the Ly-$\alpha$ forest. Our modeling of
the thermal state of the IGM is closely related to that in \citet{Furlanetto:2009kr}, except that we implement a similar general approach
using numerical simulations, which allow us to construct mock Ly-$\alpha$ forest spectra and to measure the detailed statistical properties
of these spectra. The works of \citep{Trac:2008yz,Cen:2009bg} use radiative transfer simulations to model hydrogen reionization and the
thermal history of the IGM and so these authors track
some of the underlying physics in more detail than we do here. However, our approach here is faster, simpler, and more flexible, while we believe
that it nevertheless captures many of the important processes involved.

\section{Simulations}
\label{sec:sims}

Our analysis makes use of two different types of numerical simulations. First, we use the ``semi-numeric'' scheme of
\citet{Zahn:2006sg} to model reionization; this algorithm is performed 
on top of the dark matter simulation of \citet{McQuinn:2007dy}. The \citet{McQuinn:2007dy} simulation
tracks $1024^3$ dark matter particles in a simulation volume with a co-moving sidelength of $130$ Mpc/$h$. 
Using the semi-numeric technique allows
us to capture the impact of inhomogeneities in the reionization process, while providing the flexibility to explore a range
of possible reionization models. In these models, we assume that the gas distribution closely traces the simulated dark
matter distribution. We discuss the impact of this approximation when relevant.
As we will describe,
we map between the redshift of reionization of each gas element and its temperature at high redshift using the technique
of \citet{Hui:1997dp}; this mapping depends on the density of each gas element. We then produce mock Ly-$\alpha$ forest
spectra, according to the usual ``fluctuating Gunn-Peterson approximation'' (e.g. \citealt{MiraldaEscude:1995bu,Croft:2000hs}) although here 
we additionally account for the temperature
variations from inhomogeneous reionization.

We also make use of one of the smoothed particle hydrodynamic (SPH) simulations from \citet{Lidz:2009ca}. These simulations 
were run using the code Gadget-2 \citep{2005MNRAS.364.1105S}. This simulation tracks $2\times 1024^3$ particles
(with equal numbers of dark matter and baryonic particles) in a $50$ Mpc/$h$ simulation box. In these calculations, we ignore
the inhomogeneity of the reionization process. We use these simulations to more faithfully capture the gas distribution (for
gas elements that reionize at a given time). In constructing mock Ly-$\alpha$ forest spectra from these simulations, we first
modify the simulated gas temperatures, according to various prescriptions, in order to test how sensitive the statistical properties of the
absorption are to the gas temperature.

\section{Reionization Histories}
\label{sec:reion_hist}

In an effort to explore how the thermal state of the post-reionization 
IGM depends on the reionization history of the universe, we consider several example reionization histories. 
Our aim is to consider models that result in a wide range of possible thermal histories, while broadly maintaining 
consistency with current observational constraints on reionization. 

For simplicity, we assume (as is common) that early galaxy populations produce ionizing photons at a rate
that is directly proportional
to the rate at which matter collapses into halos above some minimum mass. The minimum mass describes the 
host halo mass above which galaxies form readily; here we adopt $M_{\rm min} = 10^9 M_\odot$. We compute the collapse fraction
from the Sheth-Tormen halo mass function \citep{Sheth:1999su}.
With these assumptions, the volume averaged ionization fraction ($\avg{x_i}$) obeys the following differential equation \citep{1987ApJ...321L.107S,1999ApJ...514..648M}:
\beqa
\frac{d\avg{x_i}}{dt} = \zeta \frac{df_{\rm coll}}{dt} - \frac{\avg{x_i}}{\bar{t}_{\rm rec}}.
\label{eq:xi_ode}
\eeqa
The first term on the right hand side of the equation describes the rate at which neutral atoms are ionized, while the second
term on the right hand side accounts for ionized atoms that recombine. The recombination time ($\bar{t}_{\rm rec}$) depends on the clumpiness of the IGM,
parametrized by a ``clumping factor'', $C=\avg{n_e^2}/\avg{n_e}^2$, and the temperature of the IGM. In solving this equation -- and for this purpose only -- we assume an isothermal
IGM. We approximate the clumping factor and the temperature as independent of redshift.
Adopting the case-B recombination rate in solving this equation, a temperature of $T_0 = 2 \times 10^4$ K, and $C=3$ (see e.g. \citealt{Pawlik:2008mr,McQuinn:2011aa} for a discussion regarding plausible values of the clumping factor) gives
\beqa
\bar{t}_{\rm rec} = 0.93 {\rm Gyr} \left[\frac{3}{C}\right] \left[\frac{1+z}{7}\right]^{-3} \left[\frac{T_0}{2 \times 10^4 K}\right]^{0.7}.
\label{eq:trec}
\eeqa
Solving the differential equation, Eq. \ref{eq:xi_ode}, suffices to compute the average ionization fraction as a function of redshift, given
the minimum mass and efficiency, $\zeta$, of the ionizing sources.

In order to 
model reionization inhomogeneities, we use the ``semi-numeric'' scheme of \citet{Zahn:2006sg}, which is based on the excursion set model of reionization developed
in \citet{Furlanetto:2004nh}.
This scheme captures the tendency for halos -- and hence galaxies -- to form first in
regions that are overdense on large scales, and to reionize before more typical locations in the universe. In the simplest version of the semi-numeric
scheme, recombinations are considered only in an average sense and are treated as simply reducing the overall efficiency at which atoms are ionized. Let us denote the resulting efficiency factor as $\tilde{\zeta}(z)$ to distinguish it from the above ionizing efficiency factor $\zeta$.
As we explain subsequently, we allow this efficiency factor to be redshift dependent. We can then consider the condition that
a region of co-moving size $R$ is ionized. In the initial conditions, the mass enclosed within this co-moving region is 
$M=4 \pi R^3 \avg{\rho_M}/3$, with $\avg{\rho_M}$ denoting the average matter density per co-moving volume. The condition for this
region to be ionized is then:
\beqa
\tilde{\zeta}(z) f_{\rm coll}(> M_{\rm min} | \delta_M, M) \geq 1.
\label{eq:ion_cond}
\eeqa
In this equation $f_{\rm coll}(> M_{\rm min} | \delta_M, M)$ is the conditional collapse fraction, i.e., the fraction of mass in halos
above the minimum mass ($M_{\rm min}$) in a region of linear overdensity $\delta_M$. Here $\delta_M$ denotes the overdensity when the
linear density field is smoothed on mass scale $M$. 

In order to tabulate a reionization redshift for many grid cells across the volume
of our simulation, we smooth the density field -- linearly evolved from the initial conditions -- on a range of mass scales, starting from large scales and stepping downward until
we reach the scale of each simulation cell. For each simulation cell, and across all smoothing scales considered, we record
the highest redshift at which the ionization barrier (Eq. \ref{eq:ion_cond}) is crossed. This highest crossing redshift is considered
to be the reionization redshift, $z_r$, for the cell in question. We tabulate reionization redshifts for each 
of $512^3$ grid cells. This provides us with a reasonable model for the expected spatial variations in the redshift of 
reionization -- and the coherence scale of these inhomogeneities -- across the simulation volume. 

Note that here we approximate the excursion set model as determining the redshift at which each {\em volume element} is reionized, although
in reality mass elements move from their initial positions, and overdense regions expand less rapidly than typical locations. This approximation
is commonly made, and is reasonable given the large size of the ionized regions \citep{Furlanetto:2004nh} and the correspondingly large coherence scale of
the spatial variations in the reionization redshift.

Another ingredient we use from the \citet{McQuinn:2007dy} simulation is the evolved non-linear
dark matter density field, interpolated onto the same grid (using CIC interpolation) as the reionization redshifts. For our calculations with this simulation, we generally assume that
the gas distribution follows the simulated, gridded dark matter distribution. Note that the smoothing introduced by gridding the dark matter particles is
comparable to the Jeans smoothing scale:
the co-moving Jeans wavenumber for isothermal gas
at $10^4$ K, is $k_J = 13 h$ Mpc$^{-1}$ at $z=5$ which is comparable to the Nyquist wavenumber of the grid, $k_{\rm Nyq} = 12 h$ Mpc$^{-1}$.
More relevant, however, is the ``filtering scale'' -- essentially a time-averaged Jeans scale -- and this should be smaller than the Jeans scale by around a factor of a few \citep{Gnedin:1997td}. In any case a single global smoothing only roughly approximates the full effect of Jeans smoothing. We will return to discuss this further in \S \ref{sec:temp_measure} and \S \ref{sec:wave_inhomog}. In particular, in order to approximately 
capture the impact of small scale structure and thermal broadening in our mock quasar spectra, we will add small-scale structure using a lognormal model. Although using the gridded dark matter density field to represent the gas distribution is inadequate for making detailed mock spectra, it suffices for our model of the temperature distribution of the low density gas.

Returning to further consider the semi-numeric modeling, an important caveat is that this algorithm does not return precisely the
expected volume-averaged ionization fraction (see the Appendix of \citealt{Zahn:2006sg} for a discussion). 
Here we simply tune $\tilde{\zeta}(z)$
to produce the desired redshift evolution of the ionization fraction. 
Although this procedure is not
ideal, small adjustments to the ionizing efficiency factor have little impact on 
the size of the ionized regions at a given volume-averaged ionization fraction, $\avg{x_i}$,
and so this approach is adequate for our present purposes.

\begin{figure}
\bc
\includegraphics[width=9cm]{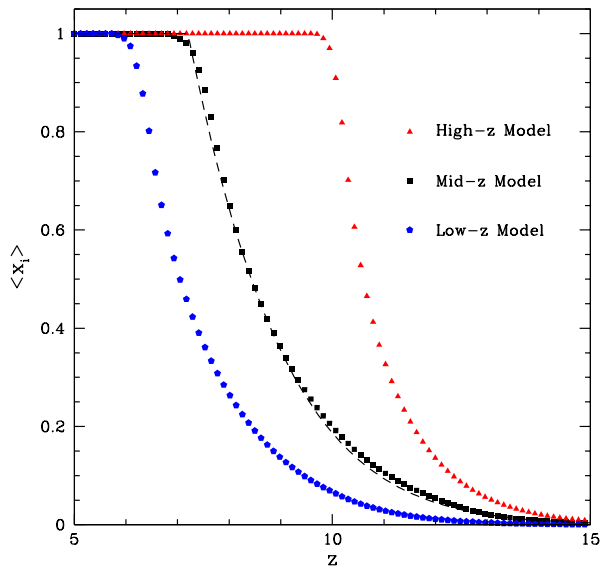}
\caption{Example reionization histories. The red triangles show the simulated volume-average ionization fraction in our semi-numeric High-z reionization model, the
black squares are for the Mid-z reionization scenario, and the blue pentagons are for a low redshift (Low-z) reionization model. The black dashed line shows the reionization
history computed by solving Eq. \ref{eq:xi_ode} with $\zeta=46$, $M_{\rm min} = 10^9 M_\odot$ and $C=3$. The semi-numeric
efficiency parameters $\tilde{\zeta}(z)$ in the Mid-z case have been tuned to match this model.}
\label{fig:xi_examp}
\ec
\end{figure}

The redshift evolution of the volume-averaged ionization fractions are shown in Fig. \ref{fig:xi_examp} for three example models. The symbols
show the average ionized fraction from the simulation cube at different redshifts. We call the three examples in the
figure the ``Low-z'' model, the ``Mid-z'' model, and the ``High-z'' model. The Mid-z model adopts a redshift dependent efficiency factor
of the form $\tilde{\zeta}(z) = 35 (1+z/13)^{1.75}$ for $z \leq 12$ and $\tilde{\zeta}=35$ for $z > 12$.
For comparison, the black dashed line shows
the solution to Eq. \ref{eq:xi_ode} for a model with $\zeta=46$, $C=3$, and $M_{\rm min} = 10^9 M_\odot$. Hence the semi-numeric scheme
in the Mid-z model has been tuned to return the ionized fraction expected from Eq. \ref{eq:xi_ode} for a plausible model.
The Low-z and High-z models are similar to the Mid-z model, except that
the efficiency factor in the semi-numeric models has been adjusted to $\tilde{\zeta}(z)=12 (1+z/11)^{0.60}$ at $z \leq 10$ and to
$\tilde{\zeta}(z)=12$ at $z > 10$ for the Low-z model and to a constant efficiency factor $\tilde{\zeta}(z)=70$ for the High-z model. (Although these 
alternative models were not themselves explicitly tuned to match particular solutions to Eq. \ref{eq:xi_ode}, the general
behavior is similar to in the Mid-z model except that reionization happens a little later/earlier in the Low-z/High-z model and
so these models also appear reasonable).

It is also useful to quantify the timing and duration of reionization, as well as the optical depth to Thomson scattering ($\tau_e$), in each model. 
Defining the ``completion'' of reionization as the redshift where the volume averaged ionization fraction first reaches $\avg{x_i} =1$,
the High-z model completes at $z=9.6$,
the Mid-z model at $z=6.7$, and the Low-z model at $z=5.8$. As one measure of the ``duration'' of reionization, we consider the redshift spread over which $\avg{x_i}$ evolves from
$\avg{x_i}=0.1$ to $\avg{x_i} = 1$. This duration is $\Delta z = 2.7, 4.3, 3.5$ for the High-z, Mid-z, and Low-z models. Note that the duration
is the longest in the Mid-z model because the ionizing efficiency factor $\tilde{\zeta}(z)$ has the strongest redshift dependence in this case. 
The electron scattering optical depths are $\tau_e = 0.088, 0.066, 0.052$ for the High-z, Mid-z, and Low-z reionization
models respectively. These values assume that the fraction of helium that is singly ionized is identical to the fraction of hydrogen that
is ionized, and ignore the slight boost expected from the free
electrons produced after HeII reionization, which we do not track in this work. 

The present constraint on $\tau_e$ from Planck CMB temperature anisotropy data \citep{Ade:2013zuv}, combined with the E-mode polarization power spectrum
at large angular scales from WMAP nine year data \citep{Bennett:2012zja}, is $\tau_e = 0.089^{+0.012}_{-0.014}$. Most of the constraining power here
comes from the WMAP polarization data. Hence our High-z model produces a $\tau_e$ close to the presently preferred value, the Mid-z model is low by $1.6-\sigma$, while the Low-z model is too low by $2.6-\sigma$. Hence our lower redshift reionization models are already marginally
disfavored, but they are still certainly worthy of further investigation. The Planck collaboration should soon announce new large-scale CMB
polarization measurements; the improved frequency coverage of the Planck satellite should help guard against foreground contamination, and
further test these models for $\tau_e$.
Although the current $\tau_e$ constraints allow higher reionization redshift models than the three
examples considered here, the $z \sim 5$ IGM temperature is insensitive to the reionization redshift
if reionization happens above $z \gtrsim 10$ (\S \ref{sec:therm_hist}, \citealt{Hui:2003hn}). While viable, we need not consider such models 
explicitly here since in these cases the $z \sim 5$ temperature will be similar to that in our High-z model.

\section{The Thermal State of the IGM}
\label{sec:therm_hist}

We now explore how the thermal state of the $z \sim 4-6$ IGM depends on the reionization history of the Universe, using
the example histories of the previous section. In this section, we focus mostly
on the Low-z and High-z models since they span a fairly wide range of possibilities for the IGM temperature at the redshifts
of interest.

The key equation describing the thermal evolution of a gas element in the IGM is (e.g. \citealt{Hui:1997dp}): 
\begin{align}
\frac{dT}{dt} =& -2 H T + \frac{2 T}{3 (1+\delta)}\frac{d\delta}{dt} + \frac{T}{\mu}\frac{d\mu}{dt} \nonumber \\
& + \frac{2 \mu m_p}{3 \rho k_B}\left({\mathcal H} - \Lambda\right).
\label{eq:tev}
\end{align}
 The first term on the right hand side accounts for adiabatic cooling owing to the overall expansion of the universe. The second 
term describes
adiabatic heating/cooling from structure formation, i.e. from gas elements contracting/expanding. In the third and fourth terms, $\mu$ is the mean mass per free particle in the gas, in units of the proton mass. The third term accounts for the temperature
change that occurs because the mean mass per particle may change with time. ${\mathcal H}$ describes the 
heating term, while $\Lambda$ is the cooling function of the gas. These terms describe the heat gain and loss per unit volume, per unit
time, in the gas. 

Let us first summarize the qualitative behavior of the solutions to Eq. \ref{eq:tev}, focusing on the low density intergalactic gas that fills most of the volume of the universe (see also \citealt{1994MNRAS.266..343M,Hui:1997dp,Hui:2003hn,Furlanetto:2009kr}). During reionization, most gas elements are rapidly ionized and change their ionization
fraction by order unity.\footnote{Sufficiently overdense regions/clumps may be only gradually ionized as reionization proceeds and the ionizing 
radiation field incident upon them grows in intensity, 
but we will neglect these, assuming that partly neutral clumps fill only a small fraction of the volume within mostly ionized regions.}
The excess energy of the ionizing photons (above the ionization threshold) goes into the
kinetic energy of the outgoing electrons, which quickly share their energy with the surrounding gas, and
raise its temperature. The first thing to consider is hence the initial temperature reached at reionization.
Provided the gas becomes highly ionized, its temperature boost during reionization
depends only on the {\em shape} of the spectrum -- and not the amplitude -- of the radiation that ionized
it. 

In detail, we expect gas elements to be ionized by radiation with a range of spectral shapes. This should be the case both
because the intrinsic ionizing spectrum will vary from galaxy to galaxy, and because
the spectral shape may be hardened by intervening absorption, which will itself vary spatially depending on the column density
of neutral gas between an ionizing source and an absorber. On the other hand, the ionized regions during hydrogen reionization
likely grow under the collective influence of numerous (yet individually faint) dwarf galaxies (e.g. \citealt{Robertson:2013bq}), and so
some of these variations may average down, provided gas elements are ionized by a combination of several sources and
the ionizing radiation arrives along various different pathways.
In any case, modeling the precise temperature input during reionization and its spatial variations requires full radiative
transfer simulations and is well beyond the scope of our approach here.
We adopt this uniform temperature boost approximation throughout, and discuss plausible values for the input temperature subsequently. Note also that in this case
the temperature boost during reionization is 
independent of density: extra heat is put into the overdense regions since more
atoms are ionized in these regions, but the heat is shared across the larger number of particles in the overdense parcel.

After a gas element is reionized, it settles into ionization equilibrium and the
UV radiation from the ionizing sources keeps the gas highly ionized (at least for the low density gas parcels that fill most
of the volume of the IGM). 
In ionization equilibrium, each recombination is balanced by a photoionization and the ionizations in turn heat
the gas; the average time between recombinations in the low density IGM is long, and so the heat input from photoionization
is significantly reduced after a parcel becomes highly ionized during reionization. In 
addition, the spectral shape of the ionizing radiation incident on
a typical gas element should soften -- i.e., the average heat input per photoionization should drop -- after reionization as the optical depth to ionizing photons decreases \citep{Abel:1999vj}.\footnote{In reality, the spectral
softening depends on how progressed reionization is {\em globally} since the hardening from absorption depends on the 
density and ionization
state of all of the gas between a source and an absorber. Here we neglect this by fixing the spectral shape incident on each
gas element after it is ionized.}

The dominant cooling processes are adiabatic cooling from the expansion of low density gas parcels and 
Compton cooling off of the CMB. As a result of cooling, although gas elements that reionize at the same time start off with identical temperatures,
irrespective of their density, parcels with differing densities will not stay at the same temperature.
In particular, the
low density elements expand and cool faster than typical regions, while overdense regions recombine faster and thus -- in ionization equilibrium -- gain 
more heat from
photoionizations after reionization. In addition, sufficiently overdense regions will be heated by adiabatic contraction.
\citet{Hui:1997dp} showed that this competition between adiabatic cooling/heating, Compton cooling, and photoionization heating,
drives the intergalactic gas to generally land on a tight temperature-density relation of the form
$T=T_0 (1 + \delta)^{\gamma-1}$. Both the temperature at mean density, $T_0$, and the slope of the temperature density relation, $\gamma$,
depend on the reionization redshift; $T_0$ falls off and $\gamma$ becomes steeper as the gas cools after reionization, until the gas
gradually loses memory of the heating during reionization. In the 
previous work of \citet{Hui:1997dp}, however, all of the gas was assumed to reionize
at the same redshift. Here we would like to generalize this to incorporate spatial variations in the redshift of
reionization (see also \citealt{Furlanetto:2009kr} and \citealt{Hui:2003hn}).

\subsection{Modeling the Thermal State}
\label{sec:temp_mod}

In general, to follow the thermal evolution in Eq. \ref{eq:tev} we should combine this equation with equations specifying the
evolution of the ionized fraction of each different particle species. However, for our present application a simpler approach should suffice.
In particular, we start by assuming that each parcel is heated to a common temperature, $T_r$, at its reionization redshift, $z_r$. We then 
follow the subsequent
thermal evolution after a gas element is ionized by assuming ionization equilibrium and that each element is highly ionized (as in \citealt{Furlanetto:2009kr}). More specifically,
we assume that both HI and HeI are highly ionized, but that HeII is not yet ionized, i.e., that HeII reionization starts
later than the high redshifts of interest
for our study. We further assume the gas is composed of only hydrogen and helium, neglecting metal line cooling, and also molecular hydrogen cooling,
which should be very good approximations for the low density IGM. In addition to adiabatic heating/cooling and Compton cooling, 
we track HI photoheating, HeI photoheating, and recombination cooling of HII/HeII using the rate expressions in the Appendix
of \citet{Hui:1997dp}. We ignore collisional ionizations, and HI/HeI/HeII line excitation cooling: neglecting these processes should
be a good approximation for the low density and highly ionized gas that fills most of the IGM volume after reionization. Furthermore, we ignore other potential heating sources such as shock heating, galactic winds, blazar heating, etc..(see e.g. \citealt{Hui:2003hn,Chang:2011bf} and references
therein for a discussion). In the Appendix
we also derive approximate solutions using linear perturbation theory (incorporating only HI photoheating, Compton cooling, and
adiabatic cooling/heating), that are useful for fast and fairly accurate estimates (see also \citealt{Hui:1997dp}).

In principle we could calculate the adiabatic expansion/contraction term (second term in Eq. \ref{eq:tev}) directly from the \citet{McQuinn:2007dy} 
simulation, at least
under the approximation that the gas distribution traces the simulated dark matter density field. Here we instead follow the approach of
\citet{Hui:1997dp} and compute this term for tracer elements assuming their density evolution obeys the Zel'dovich approximation \citep{1970A&A.....5...84Z}.
As mentioned earlier, we do, however, extend this calculation to consider gas elements with a range of different reionization redshifts. 

The basic premise here is that gas elements with identical reionization redshifts should land on a well-defined temperature-density
relation (as supported by the tests in \citealt{Hui:1997dp} and subsequent work); we can determine this relation by solving Eq. \ref{eq:tev} for many sample gas parcels.
Incorporating, however, the spread in reionization redshifts, and that the reionization redshift of each parcel may
correlate with its density, a perfect temperature-density relation will {\em not} generally be a good description. In other words,
we follow sample gas parcels to determine the mapping between the temperature-density relation at a given redshift and
the reionization redshift and temperature, i.e., this is used to determine $T_0(z|z_r,T_r)$ and $\gamma(z|z_r, T_r)$. These
mappings can then be applied to our reionization simulation to determine the temperature of any gas element, given its
reionization redshift and overdensity. To determine $T_0(z|z_r,T_r)$ and $\gamma(z|z_r,T_r)$, we follow the thermal evolution for $20,000$ tracer elements
for many different
reionization redshifts, assuming their density evolves according to the Zel'dovich approximation, and fit separate power-laws
for each $z, z_r$, and $T_r$. 

In the Zel'dovich approximation, the density field evolves according to the equation:
\beqa
1 + \delta = \frac{1}{\rm{det}[\delta_{ij} + D(t) \psi_{ij}]},
\label{eq:zeldo}
\eeqa
with $\psi_{ij}$ denoting the initial deformation tensor, and $D(t)$ denoting the linear growth factor (normalized to unity today). The density evolution of a tracer element can then be specified by the eigenvalues of the local initial deformation 
tensor.  
As in \citet{Hui:1997dp} and \citet{1995ApJ...449..476R}, we can construct realizations of the density evolution in the Zel'dovich approximation 
by randomly drawing eigenvalues of $\psi_{ij}$ from the expected probability distribution \citep{1970Afz.....6..581D}. We do this following \citet{Hui:1999ku,Bertschinger:1993zv}.

In our fiducial model, we take the temperature at reionization to be $T_r = 2 \times 10^4$ K (see \citealt{Furlanetto:2009kr,McQuinn:2012bq} for a discussion
of this choice). In calculating the photoionization heating term after a gas parcel reionizes, we assume that the specific
intensity of the ionizing radiation 
is a power-law in frequency close to the hydrogen photoionization threshold, $J(\nu) \propto \nu^{-\alpha}$ with $\alpha=1.5$.
As mentioned previously, the heat input is insensitive to the amplitude of the ionizing radiation, provided the gas is highly ionized. This spectral shape is intended
to be somewhat harder than expected for the {\em intrinsic spectrum} of the ionizing sources, since intervening absorption will harden
this spectrum \citep{1993ApJ...418...28Z,Hui:2003hn,Furlanetto:2009kr}.

\subsection{Simulated Temperature Field}
\label{sec:temp_sim}

\begin{figure}
\bc
\includegraphics[width=9cm]{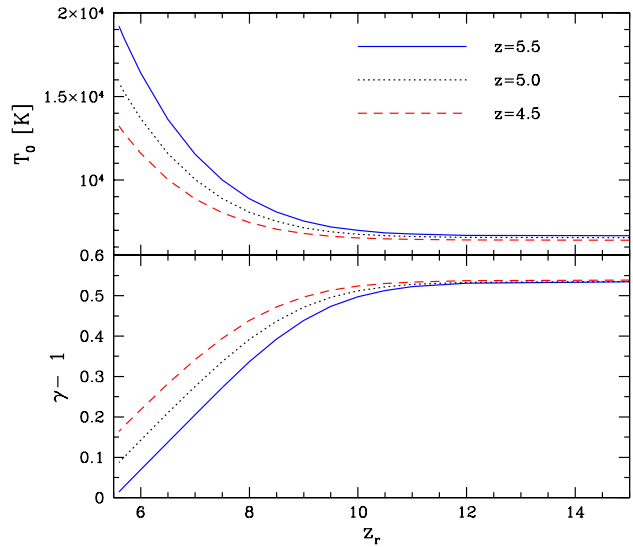}
\caption{Thermal state of gas elements with a given reionization redshift, as a function of that redshift. In each case, the gas elements are heated to
a temperature of $T_r=2 \times 10^4$ K during reionization, and the residual photo-heating after reionization is computed assuming that the
(hardened) spectral index of the ionizing sources is $\alpha=1.5$ near the HI photoionization edge. {\em Top panel:} The temperature
at mean density ($T_0$) for gas elements at each of $z=4.5,5.0$ and $5.5$ as a function of their reionization redshift. {\em Bottom panel:} This is similar to the top panel, except it shows the slope of the temperature-density relation ($\gamma-1$) rather than $T_0$. Note that although we assume that gas
elements with a given reionization redshift all land on a well defined temperature-density relation, this will not generally be
a good description once we account for the spread in reionization redshift across the universe. }
\label{fig:tden_v_zr}
\ec
\end{figure}

We now examine the properties of the simulated temperature field, modeled as described in the previous section. 
First, we consider the mapping between the temperature at mean density, $T_0$, and the slope of the temperature-density
relation, $\gamma$, for gas at various redshifts, given the reionization redshift, $z_r$, of each gas element. This
is shown, for our baseline set of assumptions, in Fig. \ref{fig:tden_v_zr} for each of $z=4.5,5.0,$ and $z=5.5$.
The values of $T_0$ are close to the temperature at reionization ($T_r = 2 \times 10^4$ K) for gas elements that ionized at redshifts
just above $z=5.5$, since these elements have had very little time to cool. On the other hand, gas parcels with higher $z_r$ have had longer to cool
and are hence at lower temperatures. For instance, gas elements that reionized at $z_r = 8$ have cooled down to $T_0 = 8,800$ K by
$z = 5.5$, more than a factor of two below the temperature at reionization. The temperatures of gas elements that reionize at sufficiently
high redshift, however, become insensitive to the precise redshift of reionization. In particular, gas elements that
reionize above $z_r \gtrsim 10$ are all at $T_0 = 6,700$ K at $z=5.5$, irrespective of $z_r$. This results mainly
because Compton cooling is very efficient at high redshift ($z \gtrsim 10$), and effectively erases the 
memory of the photoheating during
reionization \citep{Hui:1997dp}. Indeed, this is the main reason that we don't consider still higher redshift reionization
models, although they would be allowed by the present $\tau_e$ constraints as discussed in \S \ref{sec:reion_hist} (but perhaps disfavored by other data sets, see e.g. \citealt{Robertson:2013bq,Kuhlen:2012vy} for recent summaries.): 
the thermal state of the IGM is insensitive to higher redshift reionization models.

The bottom panel is similar to the top panel
except here we plot $\gamma-1$ versus $z_r$. Gas elements that reionize just above $z = 5.5$ are close to isothermal, while
elements that ionize at $z_r \gtrsim 10$ have a steeper slope, $\gamma = 1.53$. The $T_0$ and $\gamma-1$ curves at $z=4.5$ and $z=5.0$ illustrate
less sensitivity to $z_r$, since gas elements at these redshifts have had longer to cool down from their initial
temperatures at reionization. Nevertheless, the models at these lower redshifts still certainly do show some 
dependence on $z_r$. 

\begin{figure}[t]
\bc
\includegraphics[width=9cm]{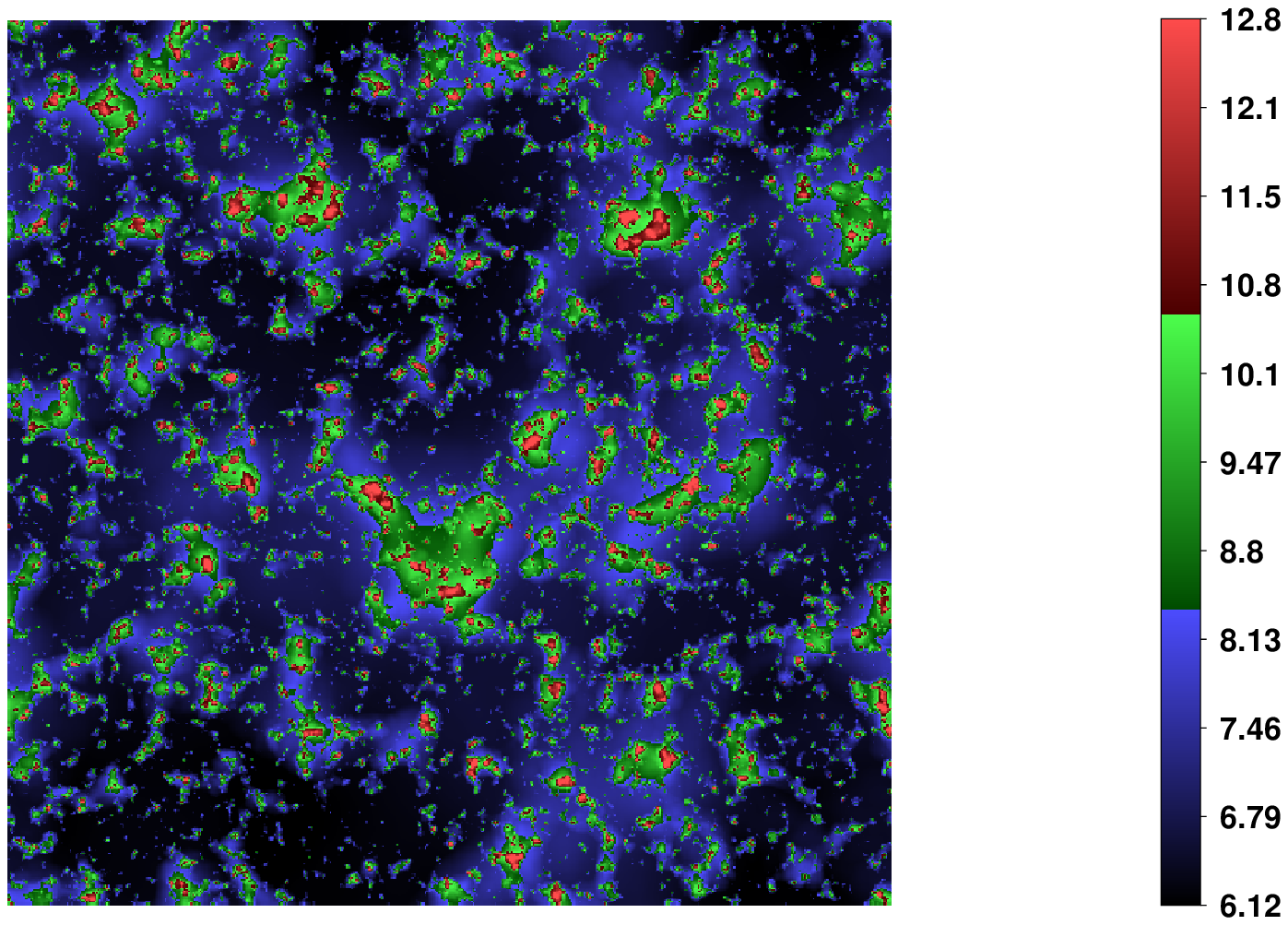}
\includegraphics[width=9cm]{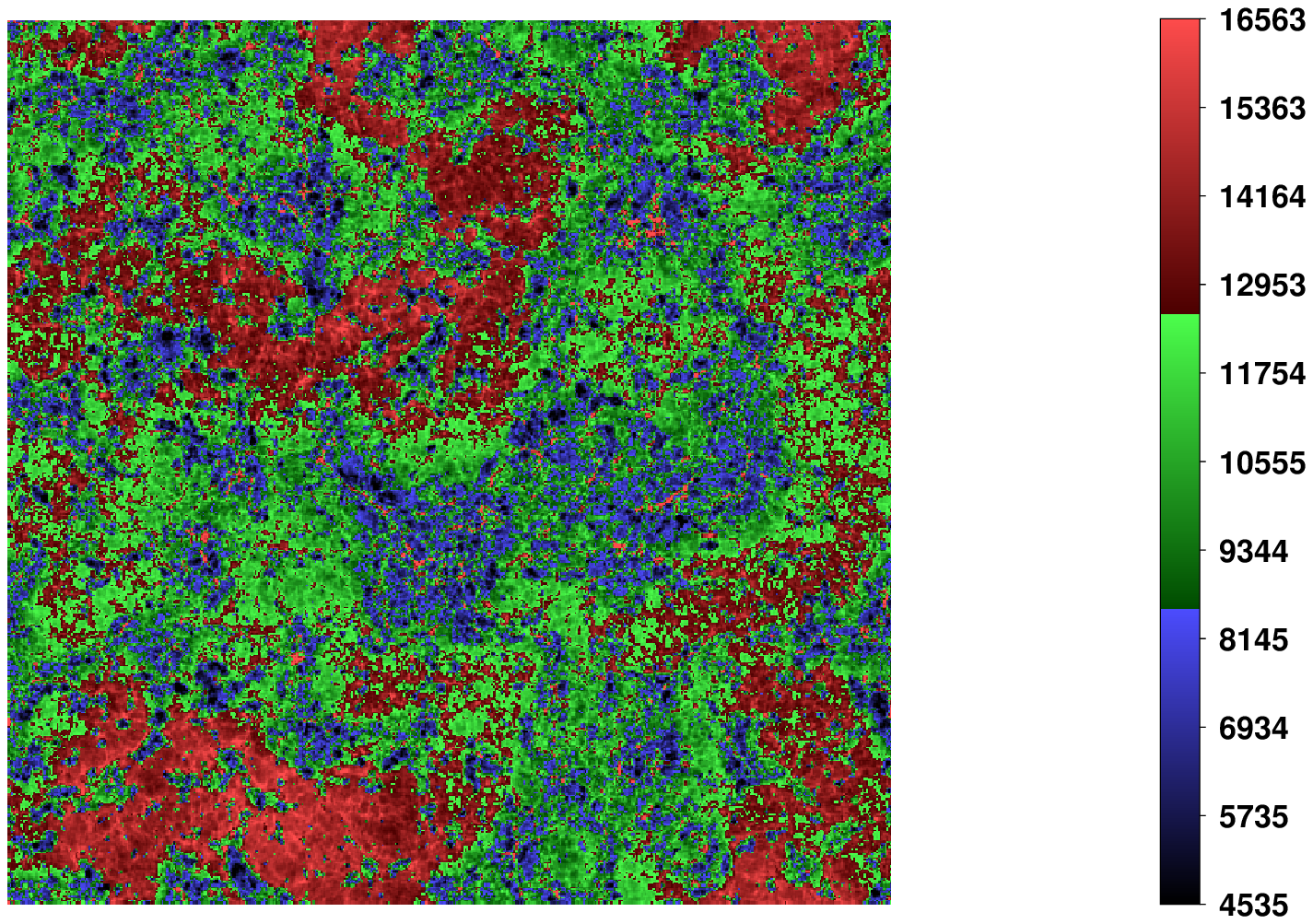}
\caption{Reionization redshifts and temperatures at $z=5.5$ in the low-z reionization model.
{\em Top panel:} The reionization redshifts for a narrow slice ($0.25$ Mpc/$h$ thick) through
the simulation. Each slice is $130$ Mpc/$h$ on a side. The red regions indicate locations with the highest reionization redshifts across
the simulation slice, while the dark regions are the last to be reionized. {\em Bottom panel:}
The temperature of the same slice as in the top panel. The red areas in this panel show the
hottest locations in the slice, and correspond to the dark regions in the top panel that
are reionized late. The dark blue regions in the temperature slice, on the other hand, are the coolest regions that reionized
first. The color scales are chosen so that $99\%$ of simulation cells in the slice shown here have redshifts and temperatures falling 
between the minimum and maximum values on the color bar.}
\label{fig:tslice_lowz}
\ec
\end{figure}

We then use the curves plotted in Fig. \ref{fig:tden_v_zr} as a mapping to predict the temperature of various grid cells in our
simulation given their overdensities, $\delta$, and reionization redshifts, $z_r$. This procedure allows us to model the temperature
field across the entire simulation volume at various redshifts. 

\begin{figure}[t]
\bc
\includegraphics[width=9cm]{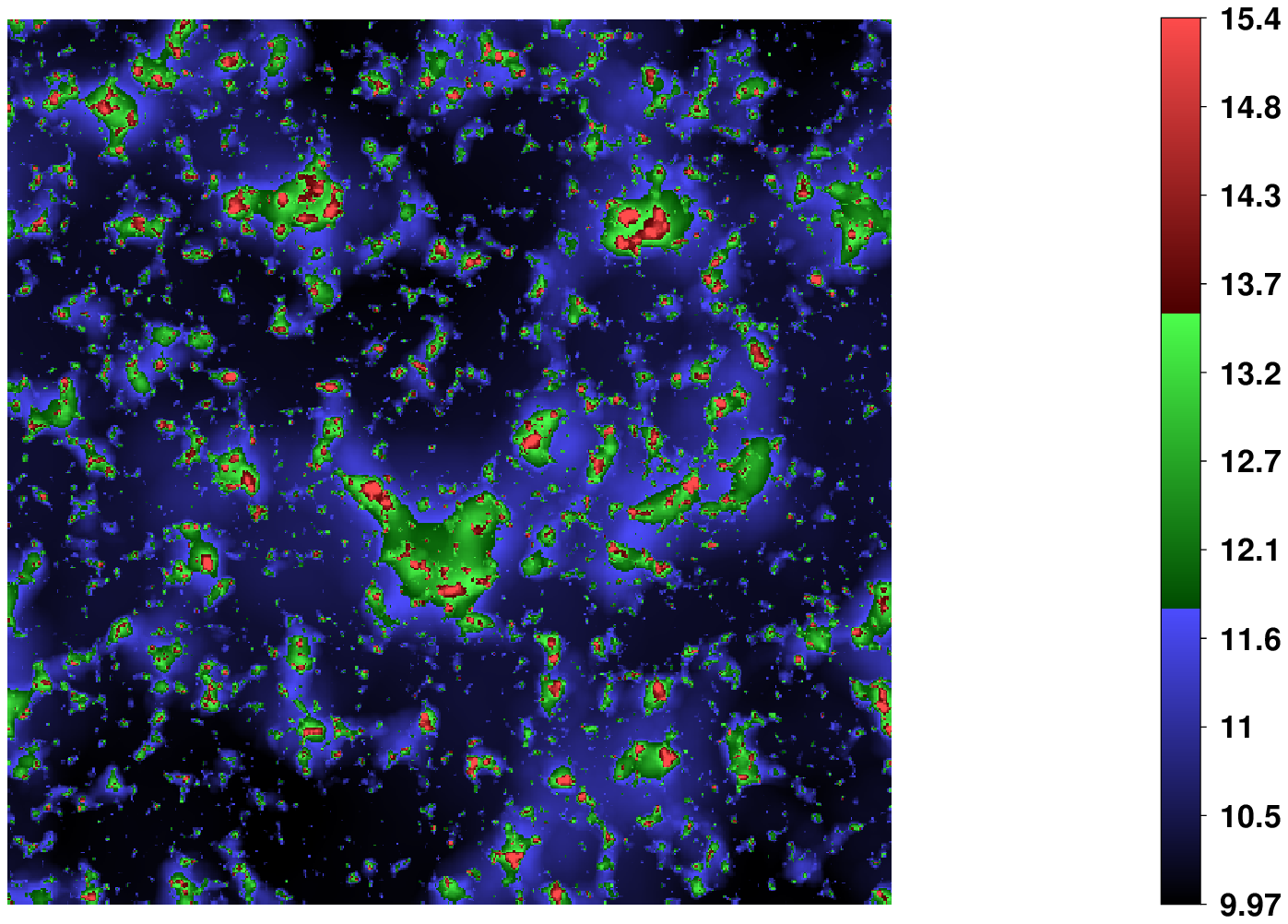}
\includegraphics[width=9cm]{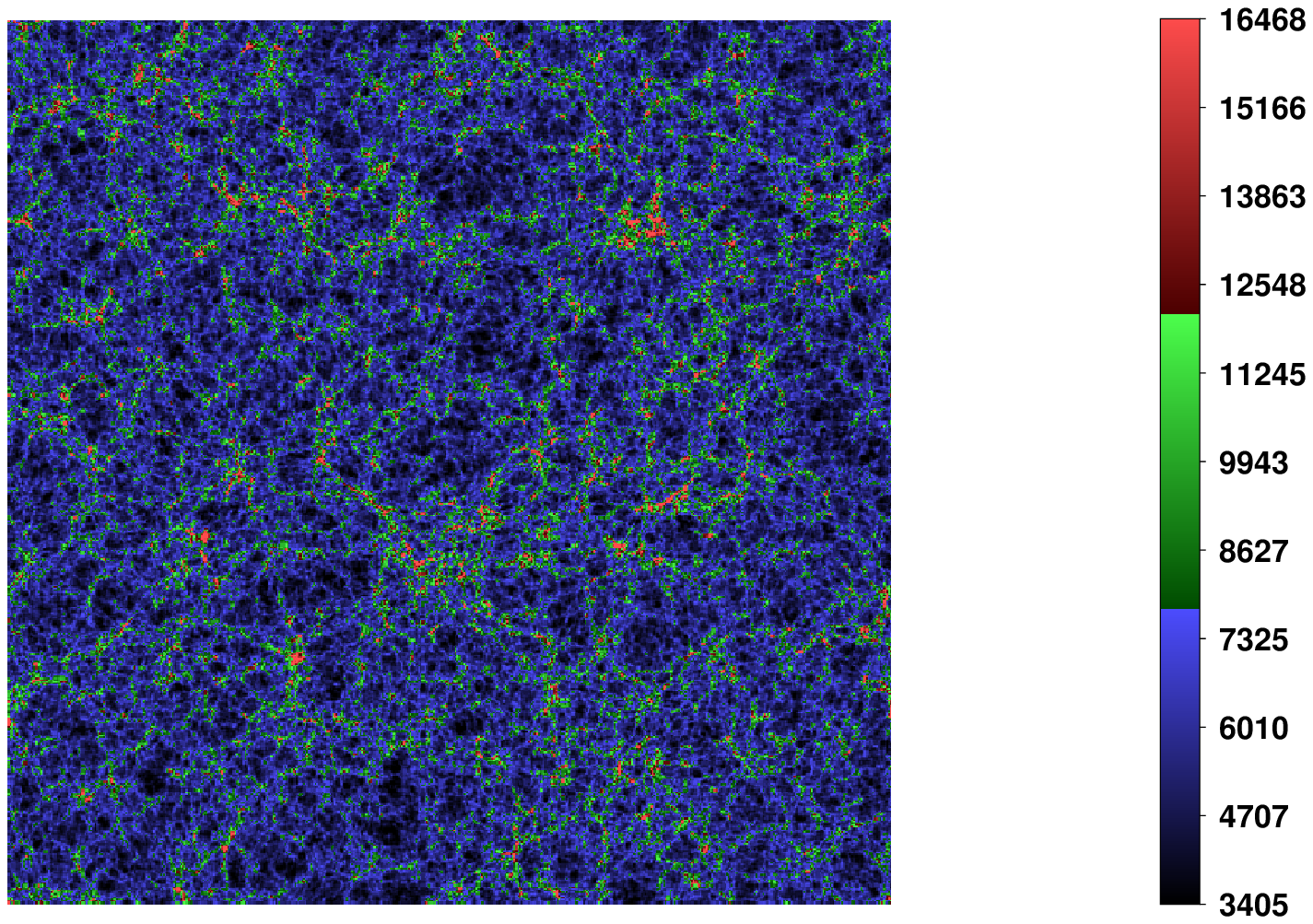}
\caption{Reionization redshifts and temperatures at $z=5.5$ in the high-z reionization model. Identical to Fig. \ref{fig:tslice_lowz},
except this figure shows the contrasting High-z model. Note that the color scale in this case also encompasses $99\%$ of the reionization
redshifts and temperatures in the simulation slice, but that these ranges are different than in the previous figure.}
\label{fig:tslice_highz}
\ec
\end{figure}

\begin{figure}
\bc
\includegraphics[width=9cm]{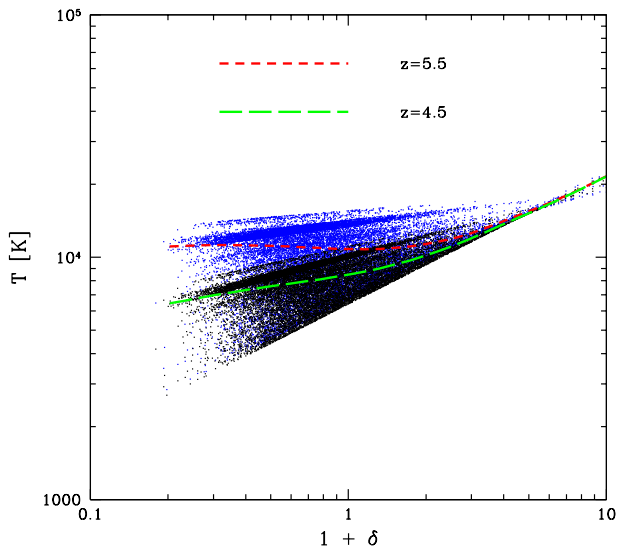}
\caption{Temperature density relations at $z = 4.5$ and $z=5.5$ in the Low-z reionization model. The
blue points show the temperature and density of gas elements from the simulation at $z=5.5$, while
the black points are the same at $z=4.5$. The red short dashed line shows the median simulated temperature
as a function of density at $z=5.5$. The green long dashed line is the same at $z=4.5$.}
\label{fig:tden_lowz}
\ec
\end{figure}

Figs. \ref{fig:tslice_lowz} and \ref{fig:tslice_highz} show the result of applying the mapping (at $z=5.5$) to the simulated density field.\footnote{In practice, we apply the mapping to the simulated density field at slightly higher redshift ($z=6.9$) since we don't currently
have outputs from this simulation at the lower redshifts of interest. Using the higher redshift output artificially reduces the variance in the density field, and the
resulting structure in the temperature field somewhat. For our present purposes, this is not important. The main effect of boosting the density
variance should be to increase the minimum and maximum density contrasts shown in scatter plots such as Fig. \ref{fig:tden_lowz}. 
Importantly, this has little impact on the median temperature-density relation and the scatter around this relation for the range of
density contrasts in our scatter plots. We have tested this explicitly using a lognormal approximation to the density field at $z=4.5$ and
$z=5.5$.} Specifically, these figures show thin slices ($0.25$ Mpc/$h$ thick) through the simulation volume, with the top
panel showing the reionization redshift and the bottom panel the corresponding temperature of cells in
the simulation volume. In the Low-z model (Fig. \ref{fig:tslice_lowz}), the temperature field
has sizable spatial variations on large scales. As anticipated earlier, these result because of the spread in the timing of reionization
across the universe. As one can infer from the slice, the regions that are at low-density (when the density field
is smoothed on large-scales) -- i.e., the ``voids'' in the density distribution -- are the last to reionize. These
regions are at the highest temperature shortly after reionization because they have
had the least amount of time to cool (see also \citealt{Trac:2008yz,Furlanetto:2009kr}).

In contrast, the temperature field in the High-z model (Fig. \ref{fig:tslice_highz}) has mostly lost memory of
the heating during reionization and so the temperature variations are more subtle here. This is expected from Figs.
\ref{fig:xi_examp} and \ref{fig:tden_v_zr}: much of the gas in this model is reionized at $z_r \gtrsim 10$,
and efficient Compton cooling mostly removes the memory of reionization in this case.
The temperature variations that are apparent in the High-z model instead reflect the usual
temperature-density relation, as the competition between cooling and heating after reionization drives overdense
regions to larger temperatures. These temperature variations 
are primarily coherent on the Jeans/filtering scale and so, as evident from the simulation slices, these fluctuations are concentrated mostly
on smaller scales than the ones induced by the spread in the timing of reionization. 

\begin{figure}
\bc
\includegraphics[width=9cm]{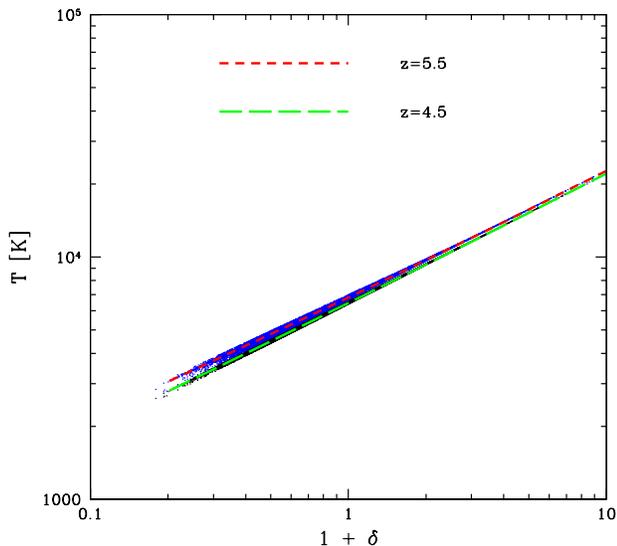}
\caption{Temperature density relations at $z = 4.5$ and $z=5.5$ in the High-z reionization model.
Identical to Fig. \ref{fig:tden_lowz}, except the results here are for the High-z reionization model.}
\label{fig:tden_hiz}
\ec
\end{figure}

A further, more quantitative description is provided by constructing 
scatter plots of the temperatures of many simulated gas elements
as a function of their densities. This is shown in Figs. \ref{fig:tden_lowz} and \ref{fig:tden_hiz} for the Low-z and
High-z reionization models, respectively. Broadly similar results may be found in earlier work by \citet{Trac:2008yz} and \citet{Furlanetto:2009kr}. The red short-dashed line in each figure shows the median gas temperature at $z=5.5$, while
the green long-dashed line is the median temperature at $z=4.5$. Fig. \ref{fig:tden_lowz}
shows that the temperature of the $z=5.5$ IGM is generally rather high -- and has a large amount of scatter at low
densities -- in the Low-z reionization model. By contrast, the temperature
in the High-z reionization model (Fig. \ref{fig:tden_hiz}) is smaller -- e.g., by 60\% for the median temperature near
the cosmic mean density  -- as is the scatter. In the Low-z model the median temperature is a fairly flat function of density
for gas less dense than the cosmic mean. 

Note that although the regions that have low density -- when the density field is averaged on large scales -- ionize last and are mostly
hotter than denser regions (see Fig. \ref{fig:tslice_lowz}), this does not fully ``invert'' the temperature-density relation. This is because the density field on the scale of the simulation grid (and at the Jeans scale) is
only somewhat correlated with the larger scale density variations that determine the spread in the timing of reionization
in our model. In any case, in agreement with previous work \citep{Trac:2008yz,Furlanetto:2009kr}, the usual temperature-density
relation is a poor description of the thermal state of the IGM in the Low-z model at $z=5.5$. 
At slightly lower redshifts, $z=4.5$, the temperature
has dropped somewhat and the scatter in the Low-z model has partially subsided, although it is still substantial. The median
temperatures in the two reionization models are closer to each other by $z=4.5$, but they still differ by 30\%. 

\begin{figure}
\bc
\includegraphics[width=9cm]{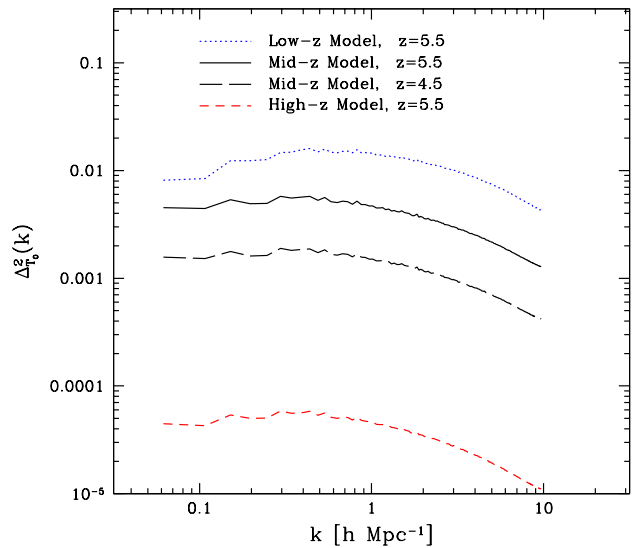}
\caption{Power spectrum of temperature fluctuations in various models. The curves show the power spectrum of 
$\delta_{T_0}(\x) = (T_0(\x) - \avg{T_0})/\avg{T_0}$ from the simulated models. The blue dotted line, the black solid line, and the red short-dashed line
are the power spectra at $z=5.5$ in the Low-z, Mid-z, and High-z models respectively. The black long-dashed line
shows the $\delta_{T_0}$ power spectrum at $z=4.5$ in the Mid-z model to illustrate how the temperature fluctuations fade with
time.}
\label{fig:power_tzero}
\ec
\end{figure}

It is also useful to calculate the power spectrum of temperature fluctuations in each model. Since we are assigning
a value of $T_0$ and $\gamma$ to each grid cell in the simulation volume, we can easily consider the power 
spectrum of $T_0$ rather
than the power spectrum of the full temperature field. The advantage of considering the power spectrum of $T_0$ is that this power
spectrum
vanishes in the case of homogeneous reionization. In the case of homogeneous reionization, the 
temperature is a power-law in the gas density, and so the full temperature field still has (mostly small scale)
fluctuations sourced directly by density inhomogeneities. Hence we consider here the power spectrum of $T_0(\x)$, or more
precisely the power spectrum of $\delta_{T_0}(\x) = (T_0(\x) - \avg{T_0})/\avg{T_0}$. Although the power spectrum of this field
is not directly observable, it nevertheless helps to characterize the temperature fluctuations from reionization.
 
The power spectra in some of our models are shown in Fig. \ref{fig:power_tzero}.
Specifically,
the curves show $\Delta^2_{T_0} = k^3 P_{T_0}(k)/(2 \pi^2)$, the contribution to the variance of $\delta_{T_0}$ per natural logarithmic
interval in $k$, i.e., per ${\rm dln}(k)$.
In the Low-z model,
the temperature fluctuations peak at a level of around $\sqrt{\Delta^2_{T_0}(k)} \approx 15\%$. 
We should keep in mind, however, that the scatter in the temperature at lower density is larger than
at mean density (see Fig. \ref{fig:tden_lowz}). As a result, the power spectra of $T_0$ shown here hence do not fully
capture the impact of inhomogeneous reionization, but they do nevertheless illustrate the spatial scale of the reionization induced
inhomogeneities as well as their redshift and model dependence.
The power spectra ($\Delta^2_{T_0}$)
are evidently fairly flat functions of $k$. This is not surprising, since the power spectra of the fluctuations in the ionization field
are also rather flat functions of $k$ during most of the EoR (e.g. \citealt{McQuinn:2006et}). At the same redshift, the $\delta_{T_0}$ power spectrum in
the Mid-z Model is $\approx 3$ times smaller in amplitude than in the Low-z model, while the amplitude of variations ($\Delta^2_{T_0}(k)$) in the High-z model are $\approx 300$ times smaller than in the Low-z model. As discussed previously, the
small fluctuations in the High-z model result because Compton cooling is efficient at high redshift and this rapid cooling effectively erases the memory of
heating at higher redshifts. Comparing the black solid and dashed lines illustrate how the fluctuations fade from $z=5.5$ to $z=4.5$ in the Mid-z model.

These models illustrate the dependence of the thermal history of the IGM on the timing of reionization; let us briefly summarize
our main findings here.
The IGM temperature for models in which
a significant fraction of the IGM volume is reionized at relatively low redshift, near $z \sim 6$, is correspondingly larger than if most of the gas is reionized
at higher redshift. In addition, the late reionization models produce sizable temperature inhomogeneities with 
fluctuations on scales as large
as $\sim$ tens of co-moving Mpc. 

\subsection{Variations around Fiducial Parameters}
\label{sec:tat_reion}

Before we proceed to discuss the observable signatures of the IGM temperature models, it is interesting to consider
how variations around our fiducial assumptions regarding the reionization temperature, $T_r$, and the shape
of the ionizing spectrum after reionization might impact the resulting thermal state of the IGM. To investigate this,
we consider models where the reionization temperature is $T_r = 3 \times 10^4$ K and $T_r=1 \times 10^4$ K to
contrast with our fiducial model in which $T_r = 2 \times 10^4$ K. The low $T_r$ model requires sources with
extremely soft ionizing spectra, and is meant to represent a lower limit to the plausible reionization temperature,
while the higher temperature $T_r = 3 \times 10^4$ K case is more reasonable (e.g. \citealt{McQuinn:2012bq}).  
In addition, for our fiducial reionization temperature
we produce models with (post reionization) spectral shapes of $\alpha=0.5$ and $\alpha=2.5$ to compare with our baseline assumption of
$\alpha=1.5$.

\begin{figure}[t]
\includegraphics[width=9cm]{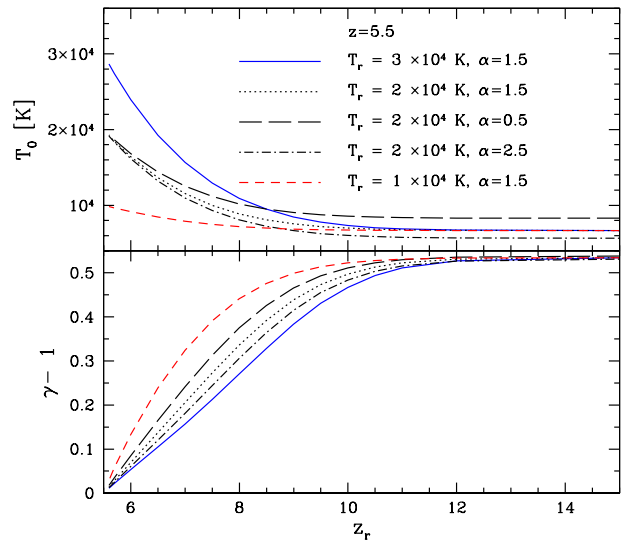}
\caption{Thermal state at $z=5.5$ for various reionization temperature and spectral shape models. This is similar to the
$z=5.5$ curves in Fig. \ref{fig:tden_v_zr}, except here we vary the reionization temperature, $T_r$, and the spectral shape,
$\alpha$. Increasing $T_r$ leads to a higher $T_0$ and a flatter $\gamma$ for recently reionized gas parcels, while parcels
that reionize at sufficiently high redshifts are insensitive to $T_r$. A harder ionizing spectrum after reionization 
(smaller $\alpha$) leads mostly to a slightly larger value of the asymptotic temperature achieved at high $z_r$.
The harder spectrum also slightly hastens the transition of $\gamma$ to its asymptotic value.}
\label{fig:temp_v_treion}
\end{figure}

First, we consider how our models for $T_0(z=5.5|z_r)$ and $\gamma(z=5.5|z_r)$ depend on the reionization temperature
and spectral shape, i.e., we regenerate the models of Fig. \ref{fig:tden_v_zr} for different values of $T_r$ and $\alpha$.
The results of these calculations are shown in Fig. \ref{fig:temp_v_treion}. The first feature to note is that $T_0$ and
$\gamma$ are independent of $z_r$ and $T_r$ in the limit of large reionization redshift: efficient cooling wipes out the memory
of the early heating history. On the other hand, the $z=5.5$ temperature is naturally quite sensitive to the reionization temperature
if reionization occurred relatively recently. One consequence of this is that the {\em scatter} in the $z \sim 5$ temperature
will be larger in low redshift reionization models for cases with larger reionization temperatures: a high reionization temperature
increases the temperature contrast between recently reionized gas parcels and those that reionized early. The increased scatter in
these models may potentially boost the observability of the temperature inhomogeneities induced by spatial variations in the timing
of reionization, as we explore subsequently.  

Another important point is that increasing $T_r$ in a high reionization redshift model will not help to mimic the $z \sim 5$ temperature
in a lower redshift reionization model, since the gas that reionized at high redshift reaches an asymptotic temperature that is insensitive
to $T_r$. On the other hand, decreasing $T_r$ (as in the $T_r = 10^4$ K curves) in a low reionization redshift model will certainly diminish the distinction between
this model and higher reionization redshift models. As we will see, however, the larger scatter and flatter trend of temperature with density
in the low $z_r$, low $T_r$ model offer potential handles for distinguishing between these models and higher reionization redshift scenarios.

Next we consider how the results vary with changes in the spectral shape, $\alpha$, as shown in the figure. These variations have a relatively minor effect. Adopting
a harder ionizing spectrum (smaller $\alpha$) after reionization increases the amount of residual late-time photoheating. This 
thereby raises
the asymptotic temperature and the asymptotic value is reached earlier. Quantitatively, the asymptotic temperature is $20\%$
higher in the $\alpha=0.5$ case than in our fiducial $\alpha=1.5$ model, and $18\%$ smaller for $\alpha=2.5$. The dependence on $\alpha$
is relatively mild compared to other uncertainties in our modeling and so we don't consider it further here.

\begin{figure}
\bc
\includegraphics[width=9cm]{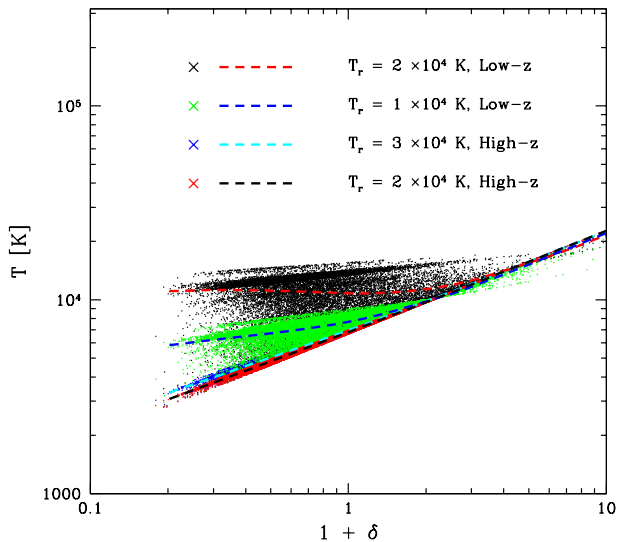}
\caption{Temperature density relation at $z=5.5$ for various reionization temperatures in the High-z and Low-z models. The ``X''s in the 
legend indicate the color of the points in the corresponding models, while the dashed lines in the same models have different colors to
promote visibility. The models in the legend are listed from top to bottom: the highest points and line (indicating the median temperature
at various densities) show the $T_r = 2 \times 10^4$ K, Low-z model; next is the $T_r = 1 \times 10^4$ K, Low-z model;
then the $T_r = 3 \times 10^4$ K, High-z model; and finally the $T_r = 3 \times 10^4$ K, High-z model.
}
\label{fig:tden_v_treion}
\ec
\end{figure}

Another perspective is to construct scatter plots in the temperature-density plane and plot median temperature-density relations for various models, as in Fig. \ref{fig:tden_lowz} and Fig. \ref{fig:tden_hiz}. This is shown for gas at $z=5.5$ in Fig. \ref{fig:tden_v_treion}. Consider first the two High-z models (the bottom two sets of points and dashed lines in the figure), which show the results of assuming $T_r = 2 \times 10^4$ K (bottom-most case with red points and a black dashed line), and $T_r = 3 \times 10^4$ K (shown as blue points and a cyan line, just above the bottom-most model). This shows that the results in this model are insensitive to $T_r$: this is as anticipated from
Fig. \ref{fig:temp_v_treion}. The next model is a Low-z case and has $T_r = 1 \times 10^4$ K (green points and blue line). This
model is clearly closer to the High-z reionization model than our fiducial Low-z case, which is shown in the figure as the upper most
black points with red line fit. However, the median temperature at low density and the scatter in the temperature 
are both larger in the Low-z, low reionization
temperature model than in the High-z models. If the scatter in the temperature, and the trend of temperature with density can be measured
observationally, this may help break the partial degeneracies between reionization redshift and temperature.

\section{Measuring the Temperature of the $z \sim 5$ IGM}
\label{sec:temp_measure}

We now turn to consider the impact of the thermal state of the IGM on the properties of the $z \sim 5$ Ly-$\alpha$ 
forest, and
on the possibility of extracting these signatures to learn about reionization. The effects of temperature
on the statistics of the Ly-$\alpha$ forest are discussed, for example, in \citet{Lidz:2009ca}. The three main effects 
are: higher temperatures produce more Doppler broadening; the recombination rate of the absorbing gas is temperature
dependent with hotter gas recombining more slowly, leading to less neutral gas and less absorption; hotter gas leads
to more Jeans smoothing, with the precise impact of this smoothing depending on the entire prior thermal history of the absorbing gas
\citep{Gnedin:1997td}. 

The enhanced Doppler broadening and Jeans smoothing in models with high temperatures each act to reduce the amount of small-scale structure in the Ly-$\alpha$ forest.
These two effects are not, however, entirely degenerate: Jeans smoothing filters the gas distribution in three dimensions, while
Doppler broadening smooths the optical depth field along the line of sight 
(e.g. \citealt{Zaldarriaga:2000mz}). Previous studies suggest that Doppler broadening impacts the small scale structure in the
forest more than Jeans smoothing, at least near $z \sim 3$ \citep{Zaldarriaga:2000mz,Peeples:2009ue,Lidz:2009ca}. At $z \sim 5$, we 
expect Jeans smoothing to have more impact, however: the high opacity in the Ly-$\alpha$ line at these redshifts implies that even slight
density enhancements can give rise to noticeable absorption lines, and these slight density variations may be erased by Jeans
smoothing. Unfortunately, it is challenging to model the impact of Jeans smoothing while incorporating a realistic model
for inhomogeneous reionization and photoheating. This requires hydrodynamic models that resolve the filtering scale, while capturing
a large enough volume to model patchy reionization. Furthermore, the filtering scale depends on the entire prior thermal history. 
In this paper, we defer this challenge to future work and assume that the effect of Jeans smoothing is sub-dominant
to that of Doppler broadening. We caution that Jeans smoothing might, however, enhance the impact of patchy reionization and 
modeling it may be necessary to robustly interpret future measurements. 

In any case, the small-scale structure in the Ly-$\alpha$
forest should be sensitive to the thermal state and the thermal history of the IGM, and so we now consider an approach for estimating the
amplitude of small-scale structure in the forest.
Here we will use the basic technique described in \citet{Lidz:2009ca}, except applied here to simulated data at higher redshift
where there is more absorption in the forest. The first issue we aim to explore here is to what extent the temperature of
the IGM is measurable at higher redshift, where the forest is significantly more absorbed. A second goal is to explore the
impact of the temperature inhomogeneities modeled in the previous section.

We briefly outline the approach of \citet{Lidz:2009ca} for measuring the small-scale structure -- and thereby extracting constraints on the 
IGM temperature -- here for completeness.
In this approach, each spectrum is convolved with a Morlet wavelet filter and
the smoothing scale of this filter is tuned to extract the amplitude of the small scale power spectrum in the forest as a function of
position across each spectrum. The Morlet filter is a plane wave, multiplied by a Gaussian and in configuration space
may be written as:
\beqa
\Psi_n(x) = K \rm{exp}(i k_0 x) \rm{exp}\left[-\frac{x^2}{2 s_n^2}\right].
\label{eq:filt_real}
\eeqa
The Fourier space counterpart, when the normalization constant $K$ is fixed so the filter has unit power (see \citealt{Lidz:2009ca}) is:
\beqa
\Psi_n(k) = \pi^{-1/4} \sqrt{\frac{2 \pi s_n}{\Delta u}} \rm{exp}\left[-\frac{(k - k_0)^2 s_n^2}{2}\right].
\label{eq:filt_four}
\eeqa
In the above equation, $\Delta u$ is the size of each spectral pixel in velocity units and $s_n$ is a suitable smoothing scale (also in velocity units) chosen
to extract the small scale power, and we set $k_0 s_n = 6$ (see \citealt{Lidz:2009ca}). Each mock spectrum is convolved with the above filter. We work with the transmission
fluctuation field, $\delta_F(x) = (F(x) - \avg{F})/\avg{F}$ where $F=e^{-\tau}$ is the transmission and $\avg{F}$ is the
ensemble-averaged mean transmitted flux. 

The transmission fluctuation, convolved with the wavelet filter, is:
\beqa
a_n(x) = \int dx^\prime \Psi_n(x - x^\prime) \delta_F(x^\prime),
\label{eq:field_filt}
\eeqa
The amplitude of this filtered field, at position ``$x$'' is given by
\beqa
A(x) = |a_n(x)|^2,
\label{eq:waveamp}
\eeqa
and characterizes the amount of small-scale structure in the transmission field. 
We generally smooth this field with a top-hat of length $L$,
\beqa
A_L(x) = \frac{1}{L} \int_{-\infty}^{\infty} dx^\prime \Theta(|x - x^\prime|; L/2) A(x^\prime),
\label{eq:smoothed_waveamp}
\eeqa
where $\Theta$ is a top-hat function. The quantity $A_L(x)$ is a measure of the average small scale power across
different portions of a quasar spectrum, and should broadly trace the temperature of corresponding regions in the IGM,
with cold regions giving a larger $A_L$ than hot regions.

\subsection{Hydrodynamic Simulations: Perfect Temperate-Density Relation Models}

As a first test, we take high redshift outputs from the hydrodynamic simulation (see \S \ref{sec:sims}) and impose temperature-density relations
before producing mock quasar spectra. This test ignores the impact of inhomogeneous reionization, but it nonetheless provides
some intuition for how well our approach can constrain the IGM temperature.

\begin{figure}
\bc
\includegraphics[width=9cm]{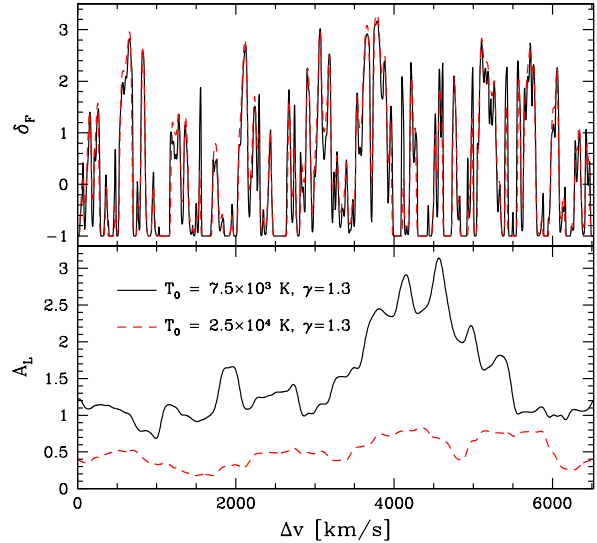}
\caption{Example sightlines and wavelet amplitudes for two different models of the IGM temperature at $z \sim 5$.
The top panel shows $\delta_F(x)$ for an example sightlines with $T_0 = 2.5 \times 10^4$ K, $\gamma=1.3$ (red dashed) and the same sightline except with $T_0 = 7.5 \times
10^3$ K, $\gamma=1.3$ (black solid). The bottom panel shows the smoothed wavelet amplitudes, $A_L$, along each spectrum.
The lower temperature model has more small scale structure and larger wavelet amplitudes. The smoothing scale $s_n=51$ km/s here, while
$\Delta u = 3.2$ km/s and $L=1,000$ km/s.}
\label{fig:examp_sightlines}
\ec
\end{figure}

Fig. \ref{fig:examp_sightlines} shows an example sightline, $50$ Mpc/$h$ in length, extracted from the hydrodynamic simulation at $z=5$
for each of two different temperature-density relation models. The top panel shows the transmission fluctuation, $\delta_F$, for
models with $\gamma=1.3$ and each of $T_0 = 7.5 \times 10^3$ K and $T_0 = 2.5 \times 10^4$ K while the bottom panel shows the
smoothed wavelet amplitudes, $A_L$, in each model. In this case, the smoothing scale $s_n$ is set to $s_n=51$ km/s, the pixel
size to $\Delta u = 3.2$ km/s, and $L=1,000$ km/s. 
In each case the intensity of the ionizing background has been renormalized
so that the global mean transmitted flux is $\avg{F}=0.20$. This is the mean transmitted flux implied by extrapolating 
the recent best-fit measurement of \citet{Becker:2012aq} to $z=5$.\footnote{Specifically, we use these authors' smooth functional fit to their measured effective optical
depth. This is an (approximate) fit to measurements in bins centered on redshifts from $z=2.15$ to $z=4.85$, and so our extrapolation of
this fit out to $z=5$ is only very slight.}

Although the differences between $\delta_F$ along the
two example sightlines are generally small, there are some noticeable differences. 
In particular, it appears that the ``spikes'' of
transmission in the colder model are more prominent. This is mostly a result of the larger Doppler widths in the hot model. 
At this redshift, the heights of the transmission spikes are often influenced
by nearby gas elements that are centered on
saturated or highly absorbed parts of the spectrum; the broad Doppler wings from this gas extend into 
adjoining unsaturated regions and thereby reduce the
height of neighboring transmission spikes. The spikes are
less impacted by the narrower Doppler wings in the colder model and remain more prominent.
Essentially, the forest has become
``inverted'' at these redshifts in comparison to at lower redshift. At sufficiently low redshift, the forest is mostly transmitted
with some prominent absorption lines interspersed. In the low redshift case most of the information about the IGM temperature comes from
narrow absorption lines. At high redshift, the forest is mostly absorbed and most of the information about the IGM temperature
is instead in the transmission spikes. 

In either case, the amount of small scale power in the Ly-$\alpha$ forest is indicative
of the temperature of the gas in the IGM. This is illustrated by the bottom panel of Fig. \ref{fig:examp_sightlines} for
the high redshift case considered here. This panel
shows the smoothed wavelet amplitudes along each line of sight. The smoothed wavelet amplitudes are larger in the cold IGM model,
with the largest differences occurring near $\Delta v \sim 4500$ km/s, close to several prominent transmission spikes in the models.

One possible complication is that long completely saturated regions will -- regardless of temperature -- have low wavelet amplitudes, $A_L$, since there
is no small scale structure in such regions. These saturated zones may in fact be {\em more prominent} in models with
low temperature since cold regions recombine more quickly, and hence have larger neutral fractions and suffer more absorption 
than hot regions.\footnote{Although the precise impact of the temperature on the wavelet amplitude PDFs shown here is not this transparent since we are comparing models at fixed
mean transmitted flux.} Fig. \ref{fig:examp_sightlines} suggests, however, that this is not a big effect at $z=5$, $\avg{F}=0.2$, although the saturated regions will be more prominent at higher redshift (see \S \ref{sec:wave_inhomog}). We can
guard against ``contamination'' from saturated regions by masking them before measuring the probability distribution
of the wavelet amplitudes, and by varying the smoothing scale $L$.

\begin{figure}
\bc
\includegraphics[width=9cm]{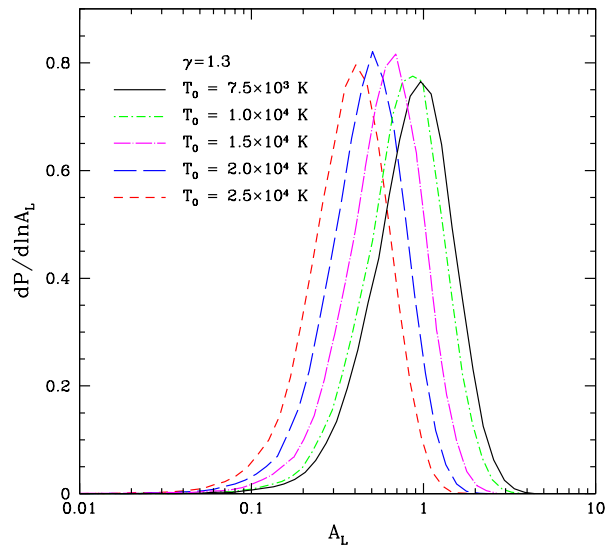}
\caption{Probability distribution of $A_L$ for various $T_0$ models at $z \sim 5$. Each model here assumes a perfect temperature
density relation with $\gamma=1.3$, and in each case the mean transmitted flux
has been fixed -- by adjusting the intensity of the ionizing background -- to $\avg{F_\alpha}=0.20$. As in Fig. \ref{fig:examp_sightlines}, 
the smoothing scale has been set to $s_n=51$ km/s, while
$\Delta u = 3.2$ km/s and $L=1,000$ km/s.}
\label{fig:pdf_amp}
\ec
\end{figure}

To characterize the variations of the wavelet amplitudes with temperature more quantitatively, we calculate the probability distribution function (PDF) of wavelet amplitudes for ensembles
of mock spectra generated from different temperature-density relation models. The results of these calculations are shown
in Fig. \ref{fig:pdf_amp}.
The PDF is quite sensitive to the temperature
at mean density in these models. For example, the location of the peak in the wavelet amplitude PDF is at an $A_L$ that is roughly
three times larger in the coldest model shown (with $T_0 = 7.5 \times 10^3$ K), compared to the hottest model considered here
($T_0 = 2.5 \times 10^4$ K). 
For the mean transmitted flux ($\avg{F}=0.20$) and $z=5$, the wavelet PDF for $s_n=51$ km/s is
sensitive mostly to densities near the cosmic mean. As a result, we find that the wavelet PDFs here depend strongly on
$T_0$, but are insensitive to $\gamma$, the slope of the temperature-density relation. It is hence important to keep in mind that our approach
for measuring the IGM temperature is only sensitive to the temperature of the IGM close to the mean density, and it is therefore not possible
to extract
the full trend of temperature with density shown in our models (e.g., Fig. \ref{fig:tden_lowz}).

Although the PDFs depend sensitively
on $T_0$, it is also clear that the wavelet amplitudes are not {\em perfect} indicators of the temperature. In the limit
that the temperature at mean density were the only quantity that determined $A_L$, these PDFs should approach delta functions
in $A_L$.
That the wavelet PDFs have some breadth is not, however, surprising: the temperature is clearly not the only quantity that
determines the small-scale structure in the forest. That said, Fig. \ref{fig:pdf_amp} looks promising and helps to motivate
further study.

\subsection{Degeneracy with the Mean Transmitted Flux}
\label{sec:degen_amf}

One other potential issue, however, is that the wavelet PDF is also sensitive to the somewhat uncertain value of the mean transmitted 
flux, $\avg{F}$. Although the present statistical uncertainties on this quantity are $\sigma_{\avg{F}}/\avg{F} \leq 10\%$ near $z \sim 5$ (\citealt{Becker:2012aq}), the systematic uncertainties are significantly larger. In particular, it is difficult
to estimate the unabsorbed quasar continuum level, especially at the redshifts of interest for this study, where the
absorption in the forest is very large (e.g. \citealt{2008ApJ...681..831F}). The {\em measurement} of the wavelet PDF itself should, however,
be fairly robust to uncertainties in the level of the unabsorbed quasar continuum. This is the case because we consider the statistics
of the transmission fluctuations, $\delta_F = (F - \avg{F})/\avg{F}$, for which a (multiplicative) error in the continuum normalization
divides out (see \citet{Lidz:2009ca} for a discussion and tests with lower redshift data).

\begin{figure}[t]
\includegraphics[width=9cm]{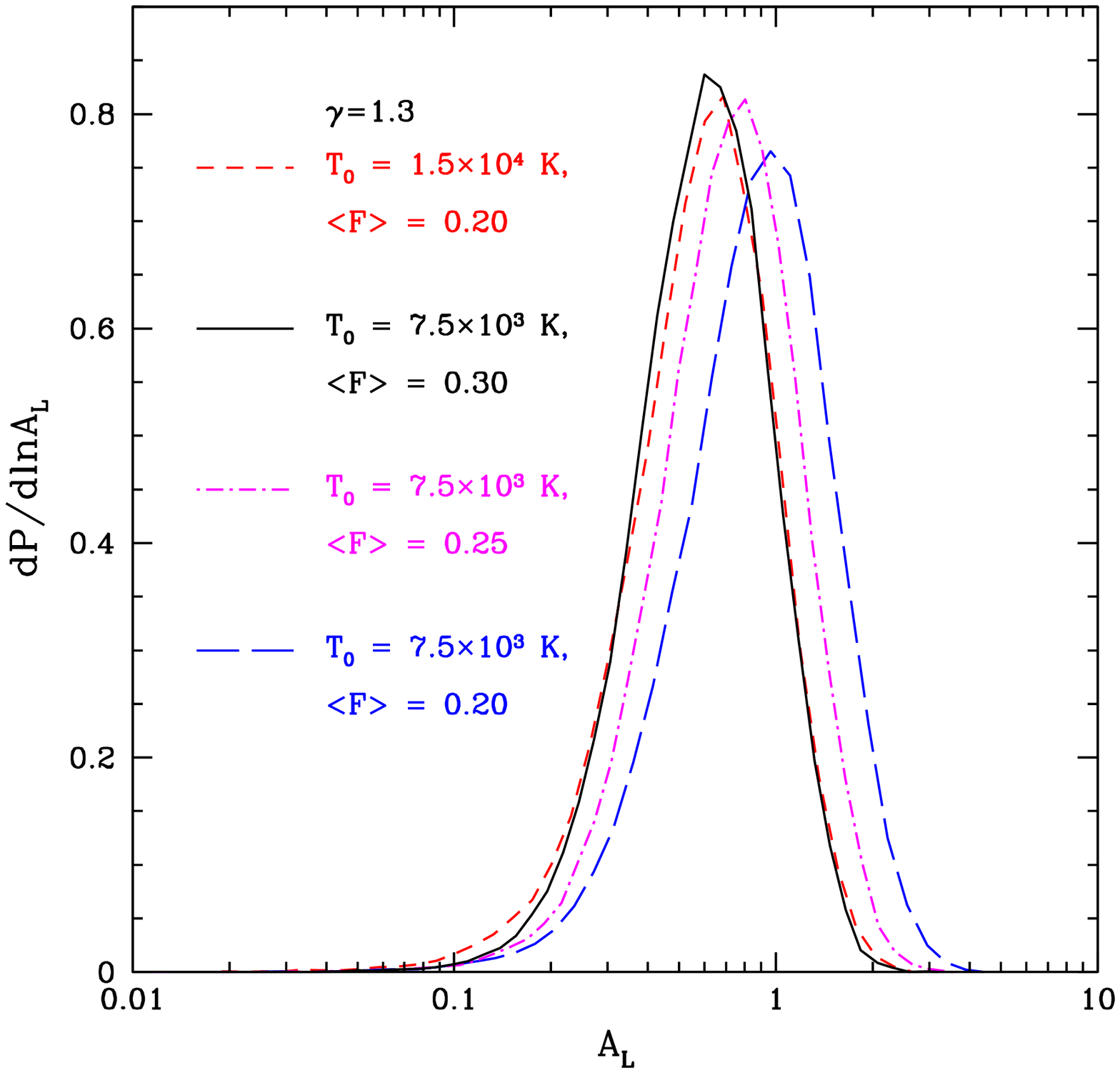}
\includegraphics[width=9cm]{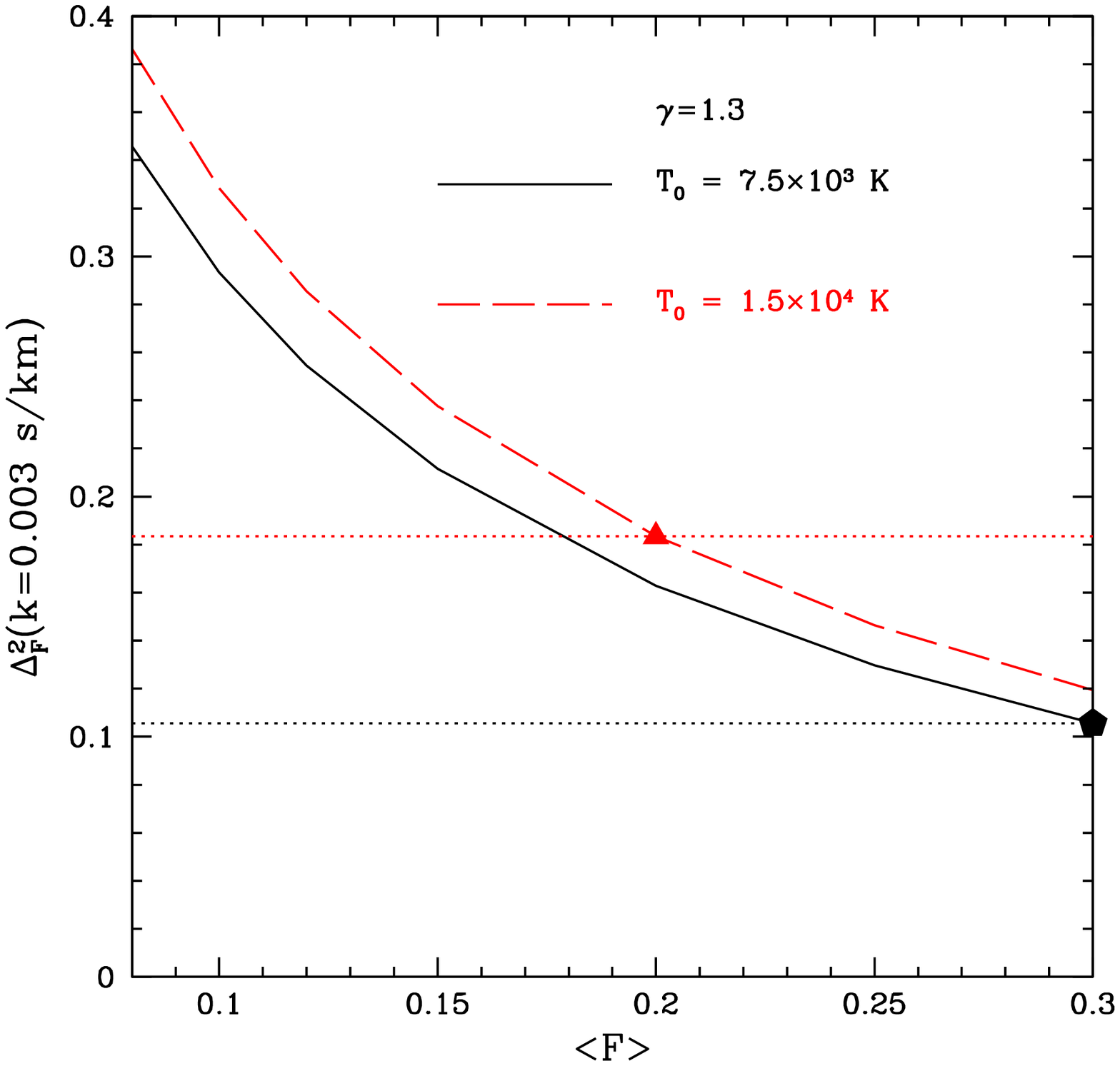}
\caption{Degeneracy with $\avg{F}$. {\it Top panel:} Although the PDF of $A_L$ is sensitive
to $T_0$, this effect is degenerate with the impact of varying $\avg{F}$. For instance, the model with $T_0 = 1.5 \times 10^4$ K
and $\avg{F} = 0.20$ is closely mimicked by a colder model with $T_0=7.5 \times 10^3$ K, yet a larger mean transmission of
$\avg{F} = 0.30$. {\it Bottom panel:} This illustrates that the degeneracy can be broken by measuring the (relatively) large
scale flux power spectrum. The curves here show the flux power spectrum, evaluated at a single convenient (larger-scale) wavenumber of $k=0.003$ s/km, 
in each $T_0$ model as a function of $\avg{F}$. The triangle and pentagon show the flux power for each model at the
$\avg{F}$ for which the wavelet amplitude PDFs are degenerate in the two models. The large scale flux power in these
two models differs appreciably and can be used to break the degeneracy. The red dotted and black dotted horizontal lines are intended 
only to guide the eye.}
\label{fig:degen_break}
\end{figure}

Nevertheless, we still need to know the mean transmitted flux very accurately: we use this measurement to in turn fix the 
intensity of the ionizing background at the redshifts of interest, which is itself quite uncertain. The amount of small
scale structure in the forest, and hence the model wavelet amplitudes, do depend on the overall mean transmitted flux. As a result,
while we should be able to measure the wavelet PDF without knowing the precise continuum normalization, our interpretation
of this measurement still requires knowing the mean transmitted flux.
To illustrate
this, we plot (in the top panel of Fig. \ref{fig:degen_break}) the $z=5$ wavelet amplitude PDF in a model with $T_0 = 1.5 \times 10^4$ K
and the preferred mean transmitted flux at this redshift ($\avg{F}=0.20$, red short-dashed line). As in Fig.  \ref{fig:pdf_amp},
the wavelet amplitudes are smaller in this model than in, for example, the cooler model with $T_0 = 7.5 \times 10^3$ K at the {\em same value of the} mean transmitted flux (blue long-dashed line in the top panel of Fig. \ref{fig:degen_break}). However, if we allow
the mean transmitted flux to increase in the colder model, the resulting wavelet PDF becomes similar to that in the hotter model. 
In particular, the black solid line shows a colder model with the mean transmitted flux increased to $\avg{F}=0.30$; this
closely matches the wavelet PDF in the hotter model at the smaller mean transmitted flux ($\avg{F}=0.20$). This particular value 
of the mean transmitted
flux, $\avg{F}=0.30$, is well outside the presently allowed range, given 
the statistical errors on current measurements. Nevertheless, the wavelet PDF clearly shows
some degeneracy between variations in $T_0$ and in $\avg{F}$. This invites further attention, especially given the systematic
concerns associated with estimating the unabsorbed continuum level.

One way to help break this degeneracy is to combine the measured wavelet PDF with a measurement of
the flux power spectrum on {\em larger scales}. This quantity is especially sensitive to the
mean transmitted flux, and the power spectrum of $\delta_F$ has the virtue -- like the wavelet PDF --
that it is insensitive to the overall normalization of the quasar continuum. On the other hand, on sufficiently large
scales,  the gentle fluctuations in the underlying quasar continuum still likely contaminate this measurement. However, there
should still be a useful range of scales where the structure in the forest dominates over that in the continuum (see e.g. \citealt{McDonald:2004eu}) and it is
these scales that we will consider to help break the $T_0-\avg{F}$ degeneracy.
 
To illustrate how the flux power measurement may help break this degeneracy, we plot the
amplitude of transmission fluctuations, $\Delta^2_F =k P_F(k)/\pi$, for a single example wavenumber ($k=0.003$ s/km in velocity units, or $k=0.39 h$ Mpc$^{-1}$
in co-moving units at $z=5$) as a function of mean transmitted flux.  The 1-D flux power spectrum ($P_F(k)$) is fairly flat on large scales and so the precise $k$ considered
here is not especially important.  The bottom panel of Fig. \ref{fig:degen_break} shows the flux power spectrum at $k=0.003$ s/km
as a function of $\avg{F}$ for the two values of $T_0$. In each case, $\Delta^2_F$ is a strong function of $\avg{F}$. The red
triangle and black pentagon show the power spectra in the $T_0 = 1.5 \times 10^4$ K and the $T_0 = 7.5 \times 10^3$ K
models respectively, for the values of the mean transmitted flux ($\avg{F}=0.20$ and $\avg{F}=0.30$) at which their wavelet
PDFs are degenerate. The large scale flux power spectra in these two models differ by the sizable factor of $1.7$. It should be
straightforward to measure the flux power spectrum on these scales to this level of accuracy, and so this measurement can help
pin down $\avg{F}$ and break the degeneracy.

One possible concern with this approach is that the precise relationship between $\Delta^2_F$ and $\avg{F}$ may be somewhat model dependent, and our inability to perfectly model the forest -- especially at high redshifts, potentially close to the EoR -- 
might lead us to draw spurious conclusions. At present, the only way to guard against this possibility is to test the goodness-of-fit
of our models for as wide a range of empirical tests as possible. Ideally, one would compare models with measurements of the
flux power across a wide range of scales (although significantly larger scales will be subject to contamination from power in
the quasar continuum), the wavelet PDF, the mean transmitted
flux, and perhaps the statistics of the Ly-$\beta$ forest as well (e.g. \citealt{Dijkstra:2003pd,Furlanetto:2009kr}).

\subsection{Wavelet Amplitude PDFs in Inhomogeneous Reionization Models}
\label{sec:wave_inhomog}

With the results of the previous section as a guide, we now turn to consider the wavelet amplitude PDFs in the more realistic inhomogeneous temperature models developed 
in \S \ref{sec:therm_hist}. In this case, we are using the dark matter simulations of \citet{McQuinn:2007dy} along with our model temperature
distributions.
Although the large volume of these simulations allows us to capture the reionization-induced inhomogeneities, they 
are not -- taken as is -- adequate
for capturing the small-scale structure in the Ly-$\alpha$ forest, which is the basic observable we aim to explore here. 

In order to make headway, we add small-scale structure to sightlines extracted from the simulation cube using the log-normal model, as in
\citet{2007ApJ...657...15K}. Briefly, we generate one-dimensional Gaussian random fields $\delta_G$ using the one-dimensional linear density power spectrum 
(scaled to the redshift of interest, and calculated after smoothing the three-dimensional linear power spectrum with a filter of the form $e^{-2 k^2/k_f^2}$ and $k_f = 30 h$ Mpc$^{-1}$ to loosely mimic the effect
of Jeans smoothing, \citealt{Gnedin:1997td}). From the Gaussian random realizations, we produce lognormal fields at high resolution using the transformation
$1 + \delta_{\rm LN} = e^{\delta_G - \sigma_G^2/2}$, where $\sigma_G^2$ is the variance of the Gaussian random field. As in \citet{2007ApJ...657...15K}
the lognormal field is modulated by the larger scale modes captured in the simulation ($\delta_{\rm sim}$) according to $1 + \delta = (1 + \delta_{\rm sim}) (1 + \delta_{\rm LN})$ with the (subscript-free) symbol $\delta$ denoting the density contrast with added small-scale structure. Similarly, using the simulated temperature in a coarse pixel (described by $T_0$ and $\gamma$), the temperature in 
a fine pixel becomes $T = T_0 (1+ \delta)^{\gamma-1}$.

The main disadvantage here is that the resulting sightlines have too much large scale structure: the lognormal field adds both large and small
scale modes to the simulation, and the simulation was not deficient in large scale power to begin with. This is partly mitigated by our using
a slightly higher redshift simulation output ($z=6.9$) than the redshift of interest. 
This simple approach is hence
imperfect, but the added small scale structure does nevertheless allow us to reliably model the impact of
thermal broadening on the resulting mock Ly-$\alpha$ forest spectra. As a test, we measure the flux power spectrum from the mock ``lognormal-enhanced'' sightlines and compare them with the flux power spectrum from mock spectra generated from the hydrodynamic simulation. The flux power spectrum from the
lognormal spectra is roughly $50\%$ larger than from the hydrodyamic simulations. However, the overall shape of the flux power is fairly well
captured in the lognormal case, and importantly, the shape of the flux power spectrum varies in a similar way in both calculations as
the temperature and thermal broadening are varied. Hence we believe that this approach suffices to capture the main impact of patchy reionization on
the small scale structure in the Ly-$\alpha$ forest. We caution, however, that a detailed comparison with upcoming measurements will certainly require
improvements here.

\begin{figure}
\bc
\includegraphics[width=9cm]{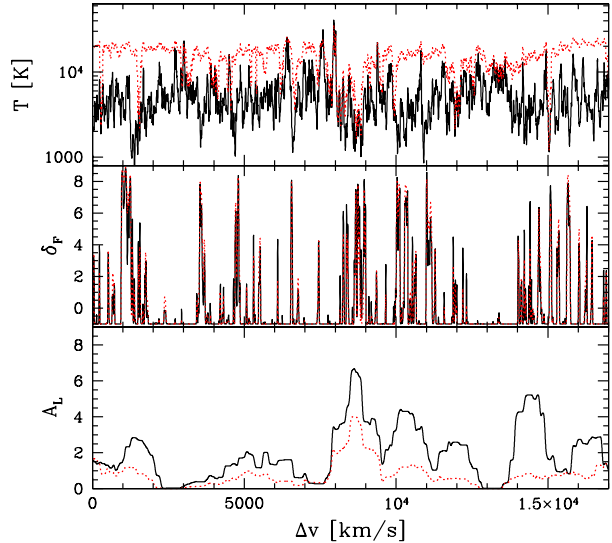}
\caption{Example sightlines and wavelet amplitudes from the Low-z and High-z reionization models. In the models here, the global mean flux
is $\avg{F}=0.1$ and $z=5.5$. In each panel the red dotted
line shows a sightline through the $T_r=3 \times 10^4$ K, Low-z reionization model while the black solid line is the same sightline,
except in this case the temperature field is drawn from the
High-z reionization model (with $T_r = 2 \times 10^4$ K). The simulated density and temperature fields have small scale structure
added according to the lognormal model, as described in the text. {\it Top panel:} The simulated temperature field. {\it Middle panel:}
The transmission field, $\delta_F$. {\it Bottom panel}: The smoothed wavelet amplitude with $L=1,000$ km/s, $s_n=34$ km/s, and
$\Delta u=2.1$ km/s. The transmission fluctuations and wavelet amplitudes are larger than in Fig. \ref{fig:examp_sightlines}, mostly because
of the lower mean transmitted flux adopted here.}
\label{fig:examp_wave}
\ec
\end{figure}

With this cautionary remark, we turn to consider the properties of mock spectra drawn from our inhomogeneous temperature models.
Fig. \ref{fig:examp_wave} shows typical example sightlines from the High-z and Low-z reionization models at $z=5.5$ and $\avg{F}=0.1$. The trends
are broadly similar to those 
in Fig. \ref{fig:examp_sightlines}: the colder models have more small scale structure than the hotter models. As a result, the transmission
field has more prominent spikes in the colder High-z model, and the smoothed wavelet amplitudes in this model (bottom panel) are 
larger than in the Low-z model. 
The inhomogeneous
models incorporate, however, the reionization-induced temperature variations that are not included in the previous model, although the impact of
these variations are generally hard to discern by eye. One can however identify that the prominent cold region in the middle of this sightline,
for example, corresponds to a pronounced peak in the smoothed wavelet amplitude field in each model. Note that the wavelet amplitudes and $\delta_F$
fluctuations are larger here than in Fig. \ref{fig:examp_sightlines} because here we consider $z=5.5$ and $\avg{F}=0.1$, while in the previous figure
we considered $z=5.0$ and $\avg{F}=0.2$. In addition, the lower transmitted flux considered here leads to more completely absorbed
regions in the mock Ly-$\alpha$ forest, and these regions have correspondingly low wavelet amplitudes. As discussed earlier, these regions do not
contain information about the IGM temperature.

\begin{figure}
\bc
\includegraphics[width=9cm]{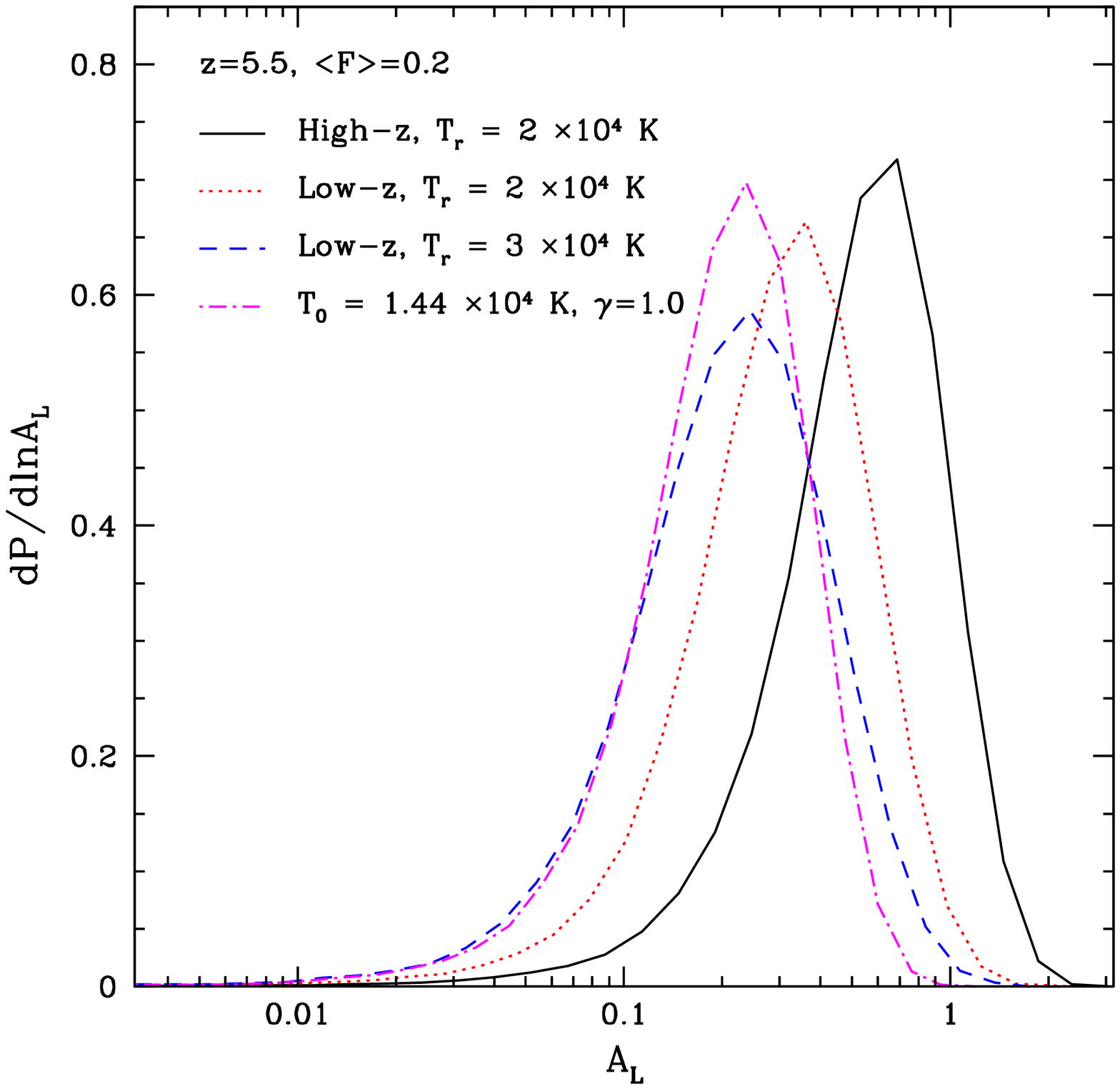}
\includegraphics[width=9cm]{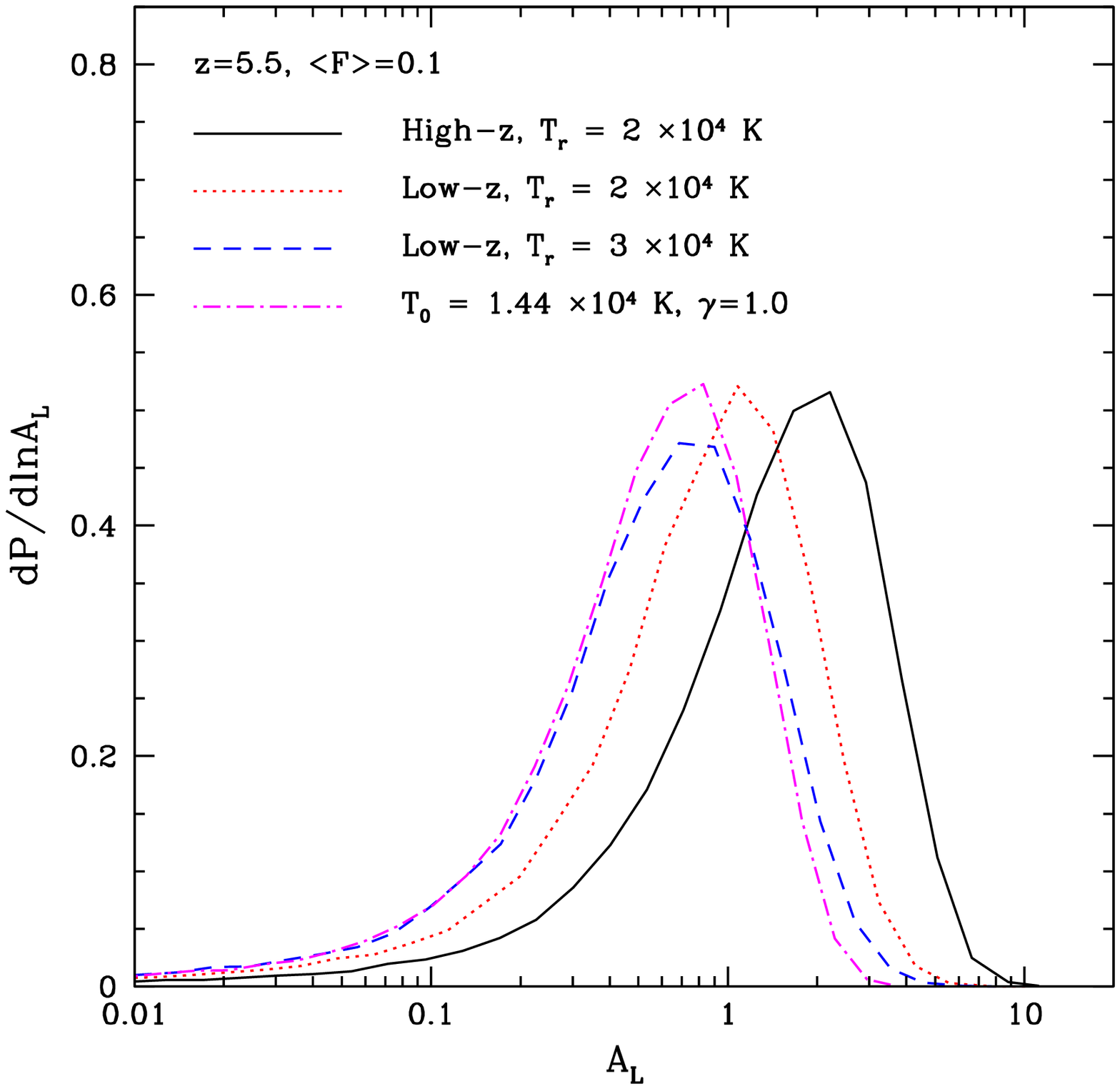}
\caption{Probability distribution of $A_L$ for various reionization and temperature models at $z=5.5$. {\it Top panel:} In this panel all models are normalized
to $\avg{F}=0.2$. The solid black curve shows the wavelet amplitudes for the High-z reionization model (with $T_r = 2 \times 10^4$ K), while the red dotted and blue dashed curves show
Low-z reionization models with reionization temperatures of $T_r = 2 \times 10^4$ K and $T_r = 3 \times 10^4$ K respectively. The magenta dot-dashed line
shows a {\em homogeneous} temperature model for comparison. In this case, the temperature was set to match the median temperature in the
 Low-z, $T_r = 3 \times 10^4$ K model for
gas at the cosmic mean density; the broader distribution in the Low-z model reflects the impact of inhomogeneous reionization. 
{\it Bottom panel:} Identical to the top panel, but here the models fix $\avg{F}=0.1$. In each case, the filter scale and pixel size are set to $s_n=34$ km/s and $\Delta u=2.1$ km/s respectively,
while $L=1,000$ km/s.}
\label{fig:pdf_al_inhomog}
\ec
\end{figure}

More quantitatively, the resulting wavelet amplitude PDFs for some example inhomogeneous temperature models are shown in Fig. \ref{fig:pdf_al_inhomog}. Since the 
mean transmitted flux at this redshift ($z=5.5$) is somewhat uncertain, we compare models normalized to each of $\avg{F}=0.2$ (top panel) and $\avg{F}=0.1$ (bottom panel) in order
to illustrate the impact of varying $\avg{F}$. Extrapolating the best fit measurement from \citep{Becker:2012aq}, we find
$\avg{F}=0.13$ at $z=5.5$ so this range should approximately bracket the expected value, although the lower end of this range
is preferred. Note that there is some evidence that the mean transmitted flux decreases more rapidly above $z \sim 5.5$ or so compared to the evolution
expected from lower redshifts (e.g. \citealt{Fan:2005es}), and so this would
further favor the lower value, and perhaps even slightly smaller numbers than considered here. Nevertheless, it is worth considering the mean transmitted
flux dependence explicitly: even if $\avg{F}=0.2$ is reached at $z=5$ rather than $z=5.5$, the temperature fluctuations could be as large at $z=5$ as in our
$z=5.5$ models if reionization is more extended than considered here. 

In each case, the wavelet filter's smoothing scales are set to $s_n = 34$ km/s and $L=1,000$ km/s, while
the pixel size is $\Delta u = 2.1$ km/s.\footnote{The pixel size here matches that of typical Keck HIRES spectral pixels. Note that the smoothing scale, $s_n$, and pixel size, $\Delta u$, here are similar, but slightly different than used
in the previous section. The large scale smoothing, $L$, is identical.} In each panel, the PDF in the High-z model is compared to Low-z models for
two different reionization temperatures, $T_r = 2 \times 10^4$ K and $T_r = 3 \times 10^4$ K. In addition, we plot the wavelet amplitude PDF for a model
with a completely homogeneous temperature field. In this isothermal case the temperature is set to $T_0 = 1.44 \times 10^4$ K, matching the median
temperature near the cosmic mean density in the Low-z, $T_r = 3 \times 10^4$ K model. Homogeneous reionization 
models with the same $T_0$ but differing
$\gamma$ would give very similar results, since the wavelet amplitudes 
are sensitive to absorbing gas near the cosmic mean density for these values of the mean transmitted flux.
Comparing the $T_0 = 1.44 \times 10^4$ K isothermal model with the Low-z, $T_r = 3 \times 10^4$ K case helps to illustrate the impact of temperature
inhomogeneities.

First, we consider how the location of the peak in the PDF varies with the reionization model. The qualitative behavior is as expected: the High-z model
peaks at the largest wavelet amplitude, while the Low-z models peak at smaller wavelet amplitude. This results because the High-z model is colder than
the Low-z models, and so it has more small-scale structure and hence higher wavelet amplitudes. The Low-z model with the higher reionization 
temperature, $T_r = 3 \times 10^4$ K, has hotter gas at the redshift of interest ($z=5.5$) than in the Low-z scenario with $T_r = 2 \times 10^4$ K
and so the PDF in the former model is peaked at smaller wavelet amplitudes. 

These trends occur in both the $\avg{F}=0.1$ and $\avg{F}=0.2$ models. In
the smaller $\avg{F}$ model the PDFs peak at larger $A_L$, reflecting the enhanced power in the $\delta_F = (F - \avg{F})/\avg{F}$ field for decreasing
mean flux. To provide a quantitative comparison, the High-z model has an average wavelet amplitude that is $73\%$ larger than in the Low-z, 
$T_r = 2 \times 10^4$ K model, while the average amplitude in the $T_r = 3 \times 10^4$ K, Low-z model is $27\%$ smaller than the $T_r =2 \times 10^4$ K 
case. These numbers are for $\avg{F}=0.2$, but the fractional differences are similar for $\avg{F}=0.1$.

Next we consider the {\em width} of the wavelet amplitude PDFs. The width arises in part because $A_L$ is not a perfect
tracer of temperature, and also because the temperature field is inhomogeneous. The latter contribution to the width 
can in principle be used to constrain the spread in the timing of reionization, and so this quantity is highly interesting. Comparing first
the inhomogeneous models in Fig. \ref{fig:pdf_al_inhomog}, it is clear that the Low-z, $T_r=3 \times 10^4$ K model has
the widest distribution of wavelet amplitudes, followed by the Low-z, $T_r=2 \times 10^4$ K model, while the High-z model
has the narrowest $A_L$ distribution of these models. This is expected, since the High-z model has the smallest temperature
fluctuations, while the Low-z, $T_r = 3 \times 10^4$ K model has the largest temperature fluctuations. It is also instructive
to compare the isothermal models (magenta dot-dashed lines in each panel) with the Low-z, $T_r=3 \times 10^4$ K models. The isothermal model has the same median temperature near the cosmic mean density as the Low-z, $T_r=3 \times 10^4$ K model. This PDF is similar, but narrower, than in the inhomogeneous temperature case. Quantitatively, the fractional width of the distribution, $\sigma_{A,L}/\avg{A_L}$, is $18\%$ larger in the Low-z, $T_r=3 \times 10^4$ K model than in the homogeneous case at $\avg{F}=0.2$ and
$11\%$ larger at $\avg{F}=0.1$. These relatively small differences seem challenging to extract, but are sufficiently interesting to merit further
investigation.

\subsection{Forecasts}

Finally, we briefly forecast the significance at which various models may be distinguished using {\it existing} data samples. Here
we will be content with rough estimates. For simplicity, we predict the expected error bar on
only the first two moments of the wavelet amplitude PDF, and compare this to the difference between some of our models. 
We consider a sample of $N_{\rm los}$ independent spectra, and assume that each spectrum has sufficient $S/N$ so that
we can estimate error bars in the sample variance limit -- i.e., we work in the limit that photon noise from the night
sky and the quasar itself, as well as instrumental noise, are negligible compared to sample variance (also termed ``cosmic variance''). Our error budget hence reflects the scatter expected -- given the large scale structure of the universe and the limited
volume probed by our hypothetical survey -- around the true value that would be obtained if we could average over an infinite
volume.

It is instructive to first consider the type of quasar spectra that are required to measure the wavelet amplitudes in
the sample variance limit. The first obvious requirement is that the spectral resolution needs to be high enough to resolve
the thermal broadening scale, which is on the order of $\sim 10$ km/s. This can be achieved with, for example, Keck HIRES spectra
which have a spectral resolution of FWHM $= 6.7$ km/s. Spectra from the MIKE spectrograph
on Magellan would partly resolve the thermal broadening scale: the resolution of these spectra is a factor of $\sim 2$ worse
than HIRES (e.g. \citealt{Becker:2010cu}). Next, we consider the impact of photon and instrumental noise. In particular,
we estimate the $S/N$ (at the continuum) per HIRES pixel at which the expected shot-noise is a small fraction of the average wavelet amplitude in plausible models. The mean wavelet amplitude from the noise should be roughly 
$\avg{A_{\rm noise}} \approx (N/S)^2/\avg{F}$ \citep{Hui:2000rw,Lidz:2009ca}. In order for the noise to be sub-dominant, we impose
that the noise should be less than $10\%$ of the
mean wavelet amplitude in our Low-z, $T_r = 2 \times 10^4$ K model at $\avg{F}=0.2$ (which has $\avg{A}=0.35$). This requires a
$S/N \gtrsim 12$ at the continuum, per $2.1$ km/s HIRES pixel. This is a fairly stringent requirement for quasars at the high
redshifts of interest for our proposed measurements, but this sensitivity has been reached already in previous work. We 
could likely make a less stringent
requirement on the $S/N$ of the data sample: this would just necessitate careful shot-noise subtraction, and boost our error budget
somewhat.   
Currently, we are aware of 
roughly $\sim 10$ HIRES spectra in the published literature at $z \sim 5-5.7$ that meet our $S/N$ and resolution criteria 
(e.g. \citealt{Becker:2011ee}). There are substantially larger numbers of lower resolution and $S/N$ spectra
from the SDSS that could potentially be followed-up at higher resolution and sensitivity to improve the statistics here. For example, there
are $36$ SDSS-DR7 quasars in the $z =[5.0,5.2]$ redshift bin of \citet{Becker:2012aq}.

We now proceed to estimate the sample variance errors.
The first quantity of interest is the sightline-to-sightline scatter in the wavelet amplitudes, averaged over the entire Ly-$\alpha$ forest
region of each quasar. We define $P_{A,L}(k)$ to be the power spectrum of the fluctuations in the wavelet amplitude after smoothing
the amplitudes on scale $L$, i.e., the power spectrum
of $\delta_{A,L}(x) = (A_L(x) - \avg{A_L})/\avg{A_L}$. We relate the expected error bars to this power spectrum, and estimate them
by measuring $P_{A,L}(k)$ from our simulated models. This approach has the advantage that we can approximately extrapolate
$P_{A,L}(k)$ to scales beyond that of our simulation box and roughly account for missing large scale Fourier modes.
This power spectrum of $\delta_{A,L}(x)$ is a four point function of the flux and is related
to the power spectrum of $A(x)$ (the {\em unsmoothed} wavelet amplitude power spectrum) from Eqs. \ref{eq:waveamp} and
\ref{eq:smoothed_waveamp} by
 \beqa
P_{A,L}(k) = \left[\frac{\rm{sin}(k L/2)}{k L/2} \right]^2 P_A(k).
\label{eq:power_filt}
\eeqa
This is just a filtered version of the wavelet amplitude power spectrum. From each independent sightline, we estimate the
moments of the wavelet amplitude PDF by further averaging over a length scale $L_{\rm spec}$, comparable to the separation
(in velocity units) between the Ly-$\alpha$ and Ly-$\beta$ emission lines from the quasar. Note the distinction between the
two smoothing scales here: $L$ is the smoothing scale over which we are studying the wavelet amplitude variations, while
$L_{\rm spec}$ is the scale over which we estimate the moments from each spectrum.

The formula for the sample variance for a single sightline is then \citep{Lidz:2009ca}:
\beqa
\frac{\sigma^2_{A,L}(L_{\rm spec})}{\avg{A}^2} = \int_{-\infty}^{\infty} \frac{dk^\prime}{2 \pi} \left[\frac{\rm{sin}(k^\prime L_{\rm spec}/2)}{k^\prime L_{\rm spec}/2} \right]^2 P_{A,L}(k^\prime). \nonumber \\
\label{eq:var_al}
\eeqa
In a sample of $N_{\rm los}$ independent Ly-$\alpha$ forest sightlines, the expected (1-$\sigma$) fractional error on $\avg{A_L}$ (in the sample variance limit) is:
\beqa
 \frac{\delta \avg{A_L}}{\avg{A_L}} = \frac{1}{\sqrt{N_{\rm los}}} \frac{\sigma_{A,L}(L_{\rm spec})}{\avg{A_L}}.
\label{eq:err_mean}
\eeqa
We can compare this estimate of the fractional error in the average wavelet amplitude with the difference between the average
amplitudes in some of the models of the previous section.

Next, we want to consider the expected error bar on the second moment of the wavelet PDF. This second moment provides one
diagnostic for the impact of temperature inhomogeneities from patchy reionization. We would like to check whether
the variance of the $A_L$ distribution is broad enough to imply patchy reionization. This requires computing the
``variance of the variance''; in particular, we want the variance of an estimate of $\sigma^2_{A,L}$ when this
quantity is estimated from a sightline of length $L_{\rm spec}$. We calculate this quantity assuming that $A_L$ approximately
obeys Gaussian statistics. In this case, one can show that the desired variance is:
\begin{align}
{\rm Var}[\sigma^2_{A,L}(L_{\rm spec})] =& 2 \avg{A_L}^4 \int_{-\infty}^{\infty} \frac{dk^\prime}{2 \pi} \left[\frac{\rm{sin}(k^\prime L_{\rm spec}/2)}{k^\prime L_{\rm spec}/2} \right]^2 \nonumber \\
& \times \int_{-\infty}^{\infty} \frac{dk^{\prime \prime}}{2 \pi} P_{A,L}(k^\prime - k^{\prime \prime}) P_{A,L}(k^{\prime \prime}) \nonumber \\
& + 4 \avg{A_L}^4 \sigma^2_{A,L}(L_{\rm spec}).
\label{eq:var_al}
\end{align}
The expected error on an estimate of $\sigma^2_{A,L}$ from a sample of $N_{\rm los}$ independent sightlines is then:
\beqa
\frac{\delta \sigma^2_{A,L}}{\sigma^2_{A,L}} = \frac{1}{\sqrt{N_{\rm los}}} \frac{\sqrt{{\rm Var}[\sigma^2_{A,L}(L_{\rm spec})]}}{\sigma^2_{A,L}}.
\label{eq:var_nlos}
\eeqa

We can now plug numbers into Eqs. \ref{eq:err_mean} and \ref{eq:var_nlos} to estimate the ability of current samples to
constrain some of our models. We assume that $N_{\rm los} = 10$ sightlines are available for our study, and take $z=5.5$, $\avg{F}=0.2$ 
here. We assume the true model is the $T_r = 3 \times 10^4$ K, Low-z case and examine at what significance other models
may be distinguished from this case. Conservatively, we assume that $L_{\rm spec} = 2.5 \times 10^4$ km/s; the velocity separation
between the Ly-$\alpha$ and Ly-$\beta$ emission lines is $5.1 \times 10^4$ km/s, and so our assumed value effectively 
masks-out half of the forest. While one will want to mask-out proximity regions, damped Ly-$\alpha$ systems, prominent metal lines, etc., our choice is certainly conservative. In this case, evaluating Eq. \ref{eq:err_mean} in our assumed true model, we find that a measurement
of $\avg{A_L}$ from $N_{\rm los} = 10$ sightlines should rule out the High-z ($T_r = 2 \times 10^4$ K) model at $32-\sigma$, and
a Low-z model with the lower reionization temperature ($T_r = 2 \times 10^4$ K) at $9-\sigma$! These forecasts are  
optimistic, since
we have assumed -- for example -- perfect knowledge of the mean transmitted flux; in practice, the mean transmitted flux 
may have to be constrained separately from the
large-scale flux power spectrum (\S \ref{sec:degen_amf}). Nonetheless, we believe the overall point is robust: existing samples should
provide interesting constraints on $\avg{A_L}$.

Significantly more challenging is to measure the variance of the $A_L$ distribution well enough to identify signatures of patchy
reionization. With the optimistic ``true'' model considered here (the Low-z, $T_r=3 \times 10^4$ K case), however, we forecast that
$N_{\rm los}=10$ spectra are sufficient to rule out the homogeneous $T=1.44 \times 10^4$ K model\footnote{Recall that the temperature in this model matches the median temperature for gas near the cosmic mean density in the Low-z, $T_r=3 \times 10^4$ K model.} at $2.4-\sigma$, based on the variance of the $A_L$ distribution alone. Note that both of these estimates assumed $\avg{F}=0.2$ and our $z=5.5$ temperature models but the
expected constraints are similar for $\avg{F}=0.1$.

\section{Conclusions}
\label{sec:conclusions}

In this work, we modeled the temperature of the IGM at $z \gtrsim 5$, incorporating the impact of spatial
variations in the timing of reionization across the universe. We contrasted the $z \sim 5$ temperature in models
where reionization completes at high redshift -- near $z=10$ -- with scenarios where reionization completes
later, near $z = 6$. In agreement with previous work \citep{Trac:2008yz,Furlanetto:2009kr}\footnote{This is also
in general agreement with still earlier work by 
\citealt{Theuns:2002yc} and \citealt{Hui:2003hn}, although these two studies did not incorporate inhomogeneities in the
timing of reionization.}, we found
that the properties of the $z=5$ temperature differ markedly between these two models. The IGM is cooler in the
early reionization model, and the usual temperature-density relation is a good description of the temperature
state in this case, while the temperature state is more complex and inhomogeneous in the late reionization scenario.

We then produced mock $z \gtrsim 5$ Ly-$\alpha$ forest spectra from our numerical models, in effort to explore
the observable implications of the IGM temperature as close as possible to hydrogen reionization. In particular,
we used the Morlet wavelet filter approach of \citet{Lidz:2009ca} to extract the small-scale structure across
each Ly-$\alpha$ forest spectrum. The small-scale structure in the forest is sensitive to the temperature of the IGM,
and the filter we use is localized in configuration space, which makes it well-suited for application in cases
where the temperature field is inhomogeneous. 

Interestingly, we found that the small-scale structure in the
forest is sensitive to the IGM temperature even when the forest is highly absorbed. In particular, the transmission
field in between absorbed regions is more spiky if the IGM is cold, compared to hotter models. Using existing
high resolution Ly-$\alpha$ forest samples, one should be able to use this difference to 
distinguish between high redshift and lower
redshift reionization models at high significance. It may, however, be necessary to combine measurements
of the small-scale structure in the forest with measurements of the larger scale flux power spectrum to help
break degeneracies with the mean transmitted flux, which is hard to estimate directly at the high redshifts of
interest for these studies. 

In addition, we considered the impact of spatial variations in the timing of reionization on the width of
the wavelet amplitude distribution. We found that these variations broaden the width of this distribution, but
that the broadening is fairly subtle. This likely results in part because the temperature variations we
are interested in are coherent on rather large scales, and aliasing -- from fluctuations in the transmission
field transverse to the line of sight -- obscures our ability to measure large scale fluctuations along
the line of sight (e.g. \citealt{McQuinn:2010mq,Lai:2005ha}). Nonetheless, we forecast that our Low-z, $T_r=3 \times 10^4$
K model can be distinguished from a homogeneous temperature model at $2-3 \sigma$ with existing samples of ten high
resolution sightlines. Larger samples could improve on this, and an analysis of the small-scale structure in the
Ly-$\beta$ forest might help as well. In this paper, we focused on the small-scale structure since it is a direct
indicator of the temperature, but another approach would be to consider instead transmission fluctuations on
large scales, especially as probed in ``3D'' measurements of the Ly-$\alpha$ forest (e.g. \citealt{McQuinn:2010mq}). This
may be possible at $z \gtrsim 4$ with DESI \citep{Levi:2013gra,McQuinn:2010mq}.

To robustly interpret future measurements, our modeling should be improved in various ways. In particular, we should incorporate
inhomogeneous Jeans smoothing effects into our modeling. This might be accomplished by, for example, 
incorporating our semi-numeric
modeling on top of a large dynamic range HPM \citep{Gnedin:1997td} simulation. These 
calculations will need to face
the competing requirements of capturing the large-scale variations in the timing of reionization, while simultaneously
resolving the filtering scale. Nevertheless, we believe that measurements of the $z \gtrsim 5$ IGM temperature should
provide a valuable handle on the reionization history of the universe.

\section*{Acknowledgments} \label{sec:ThankYou}

We thank Claude-Andr\'e Faucher-Gigu\`ere and Matt McQuinn/Suvendra Dutta for providing, respectively, the hydrodynamic and N-body 
simulations used in this analysis. We thank them, Lam Hui, Lars Hernquist, and Matias Zaldarriaga for useful
discussions about the thermal history of the IGM and its measurement from Ly-$\alpha$ forest spectra. We also thank
Matt McQuinn for helpful comments on a draft.
The authors acknowledge support from NASA grant NNX12AC97G and from the NSF through grant AST-1109156.

\bibliography{references}

\section*{Appendix: Approximate Thermal History Calculations}

Here we derive an approximate analytic formula for the thermal history of an IGM gas element using
linear perturbation theory. 
Here our derivation is quite similar to the analytic calculation in
\citet{Hui:1997dp} (their \S 3.1), except here we include Compton cooling off of the CMB, which is important for our application in which
we consider high redshift reionization and the temperature at redshifts close to reionization.

\begin{figure}[t]
\includegraphics[width=9cm]{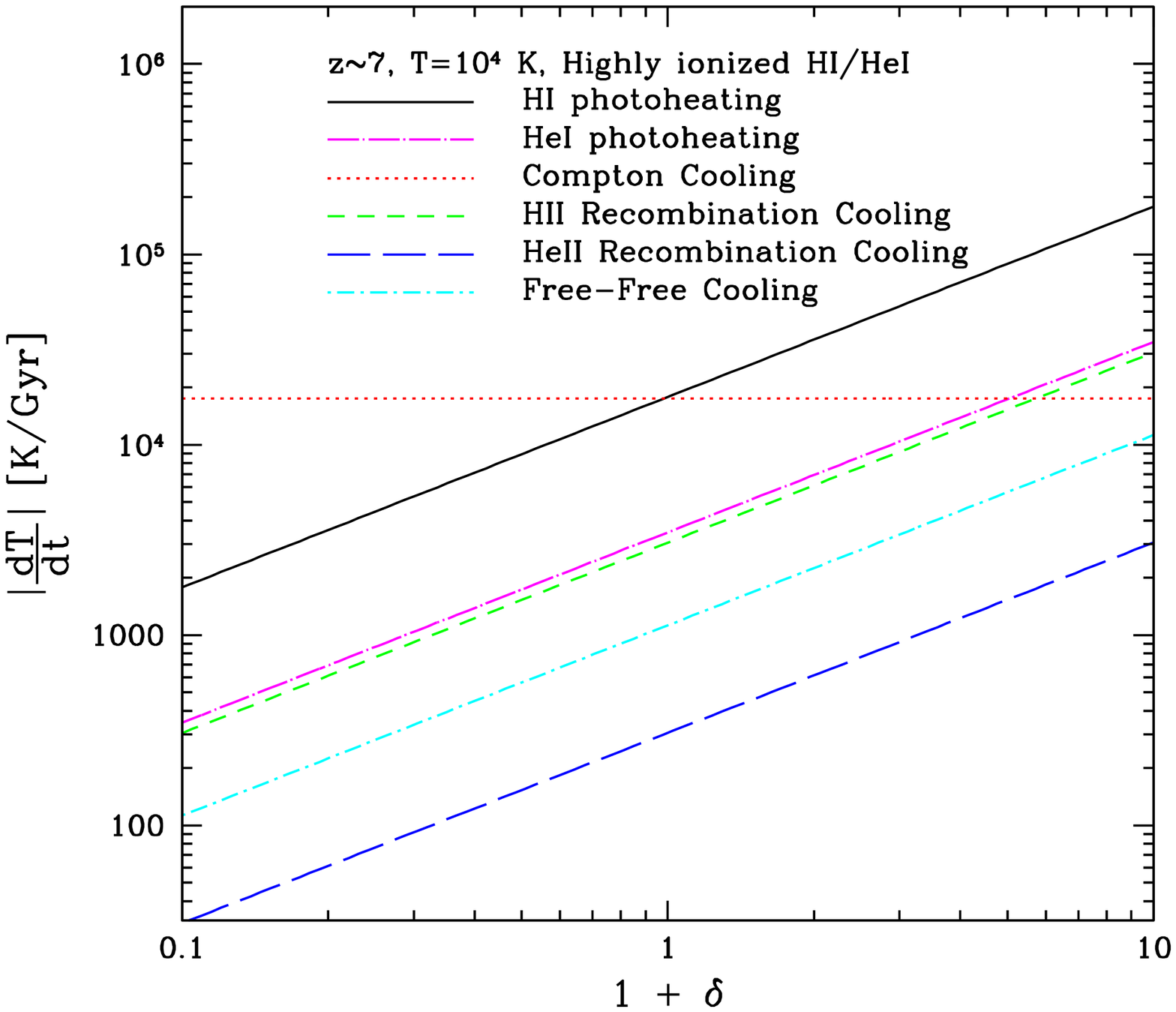}
\includegraphics[width=9cm]{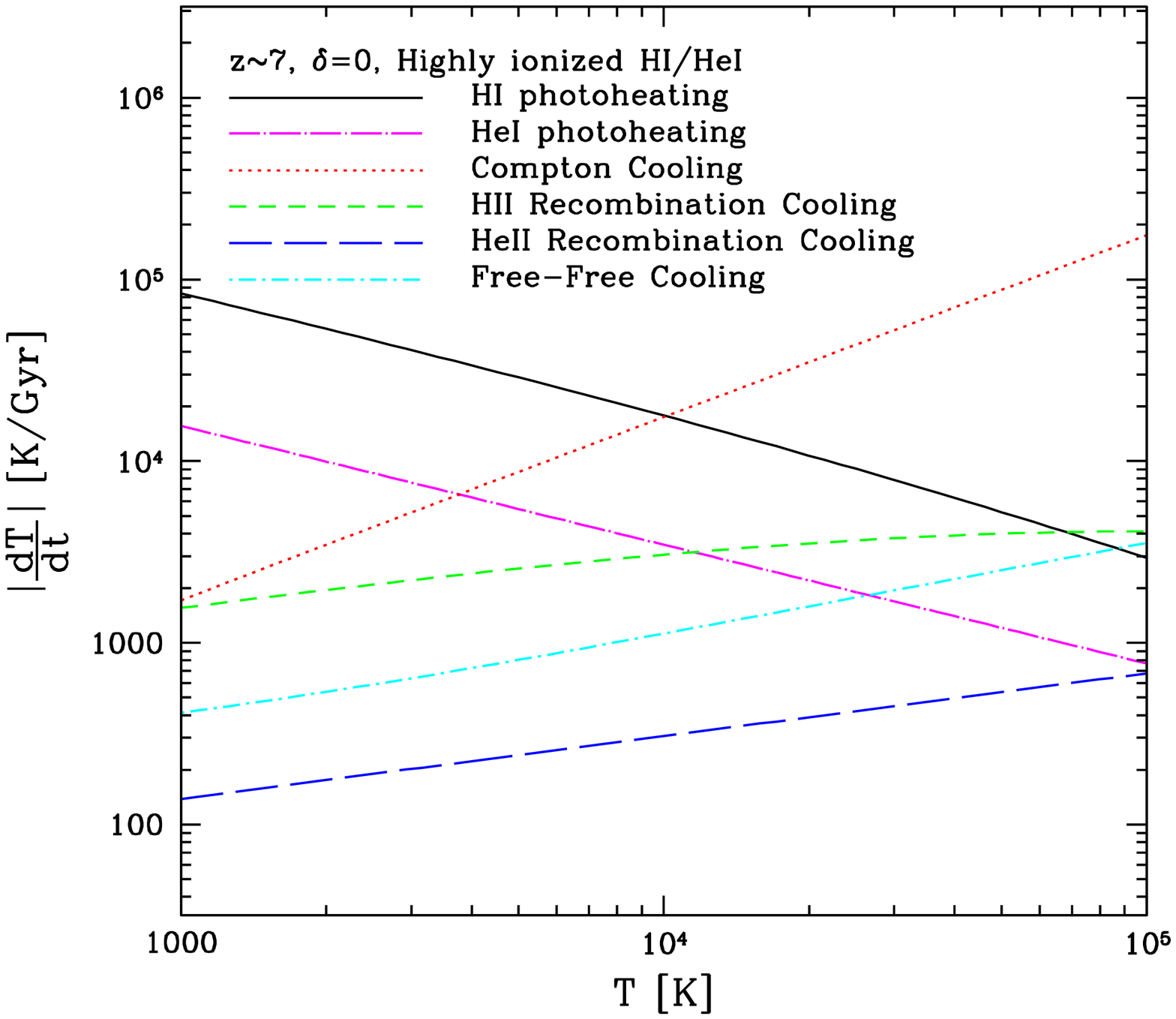}
\caption{Heating/cooling rates at $z \sim 7$. {\it Left panel}: The (absolute value of) the rates for relevant processes in the 
IGM at $T=10^4$ K as a function of density,
assuming that hydrogen is highly ionized and that helium is mostly singly-ionized. {\it Right panel}: Similar to the
left panel except the rates are shown as a function of temperature for gas at the comic mean density.}
\label{fig:temp_rates}
\end{figure}

It is instructive to first briefly examine which heating/cooling processes are important, in addition to the usual 
adiabatic heating/cooling from the contraction/expansion of gas parcels. In particular, Fig. \ref{fig:temp_rates} compares the relative importance of HI photoheating, HeI photoheating, Compton cooling,
HII recombination cooling, HeII recombination cooling, and free-free emission cooling for intergalactic gas at $z \sim 7$.
The figure assumes that the gas is in ionization equilibrium and that hydrogen is highly ionized and helium mostly singly
ionized. As in the body of the text, we are assuming that HeII reionization has not yet commenced at the redshifts of interest.
For the highly-ionized and low density intergalactic gas considered here, line excitation cooling and collisional ionizations should be unimportant.
The left hand panel considers the various heating and cooling processes for gas of fixed temperature, 
$T = 10^4$ K, as a function of density while the right hand panel shows the same for gas at the comic mean density as a function of temperature. In each
panel, the curves in the figure indicate the absolute values of the various rates, so that cooling processes are shown as positive numbers on the plot.
The photoheating
curves assume that the hardened ionizing spectrum follows a $J(\nu) \propto \nu^{-1.5}$ power-law near the photoionization 
edges.
The dominant processes are clearly HI photoheating and Compton cooling. After these processes in importance are
HeI photoheating and HII recombination cooling: these have rates that are roughly $20\%$ smaller than HI photoheating near the
cosmic mean density and $T \sim 10^4$ K at $z \sim 7$. It is also interesting to note that at the densities considered here and for
the adopted ionizing spectrum, HeI photoheating and HII recombination cooling have nearly equal magnitudes near $T \sim 10^4$ K;
since these two processes enter the thermal evolution equation with opposite signs, this leads to a partial cancellation.

As a result, a good approximation to the IGM thermal evolution (at high redshifts before HeII reionization) results
from including only adiabatic heating/cooling, Compton cooling, and HI photoheating. Note, however, that in the body of
the work we include all of the additional processes considered in Fig. \ref{fig:temp_rates}. The approximate
results here are nonetheless useful and fairly accurate, and can in turn help to build intuition.
The approximate equation for the thermal evolution is then:
\begin{align}
\frac{dT}{dt} = -2 H T + \frac{2 T}{3 (1+\delta)}\frac{d\delta}{dt} + \frac{\alpha_0 \bar{n}_e E_J}{3 (1 + \chi_{\rm He}) k_B} \left(\frac{T}{10^4 K}\right)^{-0.7} (1 + \delta) 
 + \frac{4}{3}\frac{\sigma_T a_{\rm rad} T_\gamma^4}{m_e c} \left(T_\gamma - T \right).
\label{eq:tev_approx}
\end{align}
Here $\alpha_0$ is the (case-A) recombination coefficient for hydrogen at $T=10^4$ K\footnote{We assume the case-A recombination coefficient and use the approximation $\alpha_A = 4.2 \times 10^{-13} (T/10^4 K)^{-0.7}$ cm$^3 s^{-1}$ \citep{Hui:1997dp} in this Appendix.}, $E_J$ is the average energy injected into the gas
per photoionization\footnote{Assuming a power-law spectral index, $J_\nu \propto \nu^{-\alpha}$ (with the power law accounting for hardening
from absorption), $E_J=h \nu_{HI}/(\alpha+2)$.}, $T_\gamma$ is the CMB temperature (at the scale factor of interest), $\sigma_T$ is the Thomson scattering cross
section, and $a_{\rm rad} T_\gamma^4$ is the energy
density in the CMB. The first two terms describe adiabatic cooling/heating, the third term is from photoionization heating,
and the last term accounts for Compton cooling. This equation assumes that the gas is in photoionization equilibrium, adopts an 
electron number density of $n_e = n_H + n_{He}$, and assumes the number density of free particles in the gas is $n_{\rm tot} = n_e + n_H + n_{He} =
2 (n_H + n_{He})$.

As in \citet{Hui:1997dp}, we assume a solution of the form $T = T_0 (1 + \delta)^{\gamma-1}$ and linearize ($T \approx T_0 [1+ (\gamma -1) \delta]$) to find equations
for $T_0$ and $\gamma-1$, as functions of scale factor. Let us first introduce two constants to make the notation more compact:
\beqa
{\mathcal A} = (10^4 K)^{0.7} \frac{\alpha_0 \bar{n}_e(0) E_J}{3 (1 + \chi_{\rm He}) k_B H_0 \sqrt{\Omega_m}}.
\label{eq:adef}
\eeqa
Here $\bar{n}_e(0)$ denotes the present day ($z=0$), spatially averaged, electron number density.
Note that the constant ${\mathcal A}$ has dimensions of $[{\mathcal A}] = [K]^{1.7}$. In our fiducial model with $\alpha=1.5$, the
numerical value of ${\mathcal A}$ is 
${\mathcal A} = 5.77 \times 10^5 K^{1.7}$. Second, we introduce
\beqa
{\mathcal B} = \frac{1}{H_0 \sqrt{\Omega_m} t_{\rm Comp}(0)}; \quad
t_{\rm Comp} = \frac{3 m_e c}{4 \sigma_T a_{\rm rad} T_\gamma^4}.
\label{eq:bdef}
\eeqa
Here $t_{\rm comp}(0)$ a characteristic timescale for Compton cooling today ($z=0$); this timescale falls off towards high redshift as 
$t_{\rm comp} \propto a^4$.  In our assumed cosmology, the numerical value of this constant is ${\mathcal B} = 1.15 \times 10^{-2}$.

Using the high redshift approximation for the Hubble parameter, $H \approx H_0 \sqrt{\Omega_m} a^{-3/2}$, and setting $\delta=0$ to
find an equation for $T_0(a)$ valid in linear theory, Eq. \ref{eq:tev_approx} gives the following equation for $T_0$:
\beqa
\frac{d(a^2 T_0)}{da} = {\mathcal A} a^{0.9} (a^2 T_0)^{-0.7} - {\mathcal B} a^{-7/2} (a^2 T_0) + {\mathcal B} T_\gamma(0) a^{-5/2}.
\label{eq:tzero_eq}
\eeqa
A similar equation follows for $\gamma-1$ (again valid to linear order in $\delta$, and with the approximations to the thermal
evolution equation in Eq. \ref{eq:tev_approx}):
\beqa
\frac{d(\gamma-1)}{da} = \left[\frac{2}{3} - (\gamma-1)\right] \frac{1}{a} + {\mathcal A} a^{0.9} (a^2 T_0)^{-1.7} \left[1 - 1.7 (\gamma-1) \right]
- \frac{{\mathcal B} T_\gamma(0)}{a^2 T_0} a^{-5/2} (\gamma-1).
\label{eq:gamma_eos}
\eeqa
We can safely neglect the third term in Eq. \ref{eq:tzero_eq}, and we find a solution for $T_0(a)$ of the form (with the initial condition
that the gas element is ionized at scale factor $a_r$ to a temperature $T_r$):
\beqa
u = u_r {\rm exp}\left[0.68 {\mathcal B} a^{-5/2} - 0.68 {\mathcal B} a_r^{-5/2} \right] +
0.68 {\mathcal A} (0.68 {\mathcal B})^{0.76} e^{0.68 {\mathcal B} a^{-5/2}} \int_t^{t_r} dt' t'^{-1.76} e^{-t'}.  \nonumber \\
u=(a^2 T_0)^{1.7}; \quad u_r=(a_r^2 T_r)^{1.7}; \quad t=0.68 {\mathcal B} a^{-5/2}; \quad t_r = 0.68 {\mathcal B} a_r^{-5/2}.
\label{eq:usol}
\eeqa

The corresponding solution to Eq. \ref{eq:gamma_eos} for the evolution of $\gamma$ does not have a simple closed analytic
form, but the equations can be solved numerically. 
Comparing the solutions for $T_0(z)$ from Eq. \ref{eq:usol} and $\gamma(z)$
from  Eq. \ref{eq:gamma_eos} with results given in
the body of the text, we find that the approximate solutions are good to better than $10\%$ accuracy. This accuracy is, in fact, somewhat better than might be expected given the approximations made, and may reflect cancelations between some of the neglected
terms (such as the compensating omissions of HeI photoheating and HII recombination cooling, as highlighted in Fig. \ref{fig:temp_rates}). Nevertheless, the approximate solutions seem quite useful and so we include them here.

\end{document}